\documentclass[%
 reprint, nofootinbib, amsmath,amssymb, aps,floatfix,
]{revtex4-2}
\usepackage{gensymb}
\usepackage{textcomp}
\usepackage{lipsum}
\usepackage{graphicx}
\usepackage{dcolumn}
\usepackage{bm}
\usepackage{siunitx}
\DeclareSIUnit\gauss{G}
\DeclareSIUnit\erg{erg}

\usepackage{cancel}
\usepackage{tabularx}
\usepackage{tikz}
\usepackage{amssymb}
\usepackage{amsmath}
\usepackage{relsize}
\usepackage{listings}
\usepackage{commath}
\usepackage{enumitem}
\usepackage{xfrac}
\usepackage{float}
\usepackage{rotating}
\usepackage{booktabs}
\usepackage{makecell}
\usepackage{mathtools}
\usepackage{caption}
\usepackage{subcaption}
\usepackage{multirow}
\usepackage[table,xcdraw]{xcolor}
\usepackage[version=4]{mhchem}
\usepackage[colorlinks,bookmarks=false,citecolor=red,linkcolor=blue,urlcolor=blue]{hyperref}

\begin{document}

\preprint{APS/123-QED}
\title{Decoherence-controlled collective criticality in a two-dimensional quantum Stag Hunt}
\author{Rajdeep Tah}
\email{rajdeep.phys@gmail.com}

\affiliation{Department of Physics, Florida State University, Tallahassee, Florida 32306, USA}

\begin{abstract}
    Physical decoherence can preserve the microscopic strategic neutrality (i.e. zero effective field and equal potential of the two homogeneous strategic orientations) of a quantum game while changing the thermodynamic regime of the corresponding interacting population. We demonstrate this for an Eisert--Wilkens--Lewenstein (\textit{EWL}) Stag Hunt embedded as independent nearest neighbor encounters on a square lattice. For the restricted strategies $\mathsf{Q}=i\mathbb{Z}$ and $\mathsf{D}=i\mathbb{Y}$, noisy two-player payoff matrices are determined for phase damping, depolarization, and amplitude damping and mapped exactly to channel dependent Ising parameters $\mathfrak{J}(\Gamma,p)$ and $\mathfrak{H}(\Gamma,p)$. Phase damping and depolarization show the clearest contrast: they share the same microscopic neutrality branch $\mathfrak{H}=0$, while only depolarization suppresses the interaction as $(1-p)^2$. At $\beta=1$, this produces an exact depolarization-driven square-lattice critical point at $p_{*}\approx 0.233460\ldots$, whereas phase damping remains in the ordered coexistence regime along the same neutrality branch. Monte Carlo finite size scaling is consistent with two-dimensional Ising criticality and distinguishes field driven coexistence below $p_{*}$ from a smooth crossover above it. Amplitude damping additionally reveals a strong dependence on channel placement: the post-strategy neutrality branch reaches $\Gamma=0$ at $p=1/3$ and then disappears. Resource negativity further shows that microscopic two-qubit entanglement and collective interaction strength are distinct quantities. The resulting extended lattice remains an ordinary classical Ising system.
\end{abstract}

\maketitle

\section{\label{introsec}Introduction}
Game theory and statistical mechanics are often used together to study large \textit{populations} of interacting players with two available strategies. This connection has been extensively studied via the Gibbs formulation, potential games, and thermodynamic descriptions of evolutionary games \cite{ref1, ref2, ref3, ref4}. In stochastic-type models, the randomness in the players' decisions corresponds to the thermal fluctuations, whereas the strategic potential corresponds to the energy of a particular spin configuration. Blume developed a statistical-mechanical description of strategic interactions along these lines \cite{ref2}, while exact potential games provide an accurate framework in which a player's payoff change under a unilateral strategy change is exactly reproduced by a change in the scalar potential \cite{ref1}. In lattices and graphs, these connections have led to evolutionary and equilibrium descriptions of spatial games, including explicit mappings of $2\times 2$ games to Ising-like Hamiltonians \cite{ref3, ref5, ref6, ref7}. These kinds of approaches are significantly different from standard \textit{imitation}/\textit{replicator dynamics}: in a Gibbs-type model, the game potential forms the basis for the equilibrium distribution, without requiring a particular evolutionary update rule \cite{ref2, ref7}.

A related line of work has used Ising mappings to study a variety of social dilemmas in the thermodynamic limit. Within this framework, techniques such as Nash-equilibrium mapping (NEM), susceptibility-like observables, and numerical agent-based implementations have been developed to characterize collective strategic behavior \cite{ref8, ref9, ref10, ref11}. Quantum extensions of this thermodynamic framework have also been explored. These include a one-dimensional (1D) thermodynamic analysis of the quantum Prisoner’s Dilemma \cite{ref12}, locally entangled many-body constructions \cite{ref13}, and numerical studies combining NEM with agent-based modeling \cite{ref14}. These works demonstrate that quantum strategic interactions can be incorporated into statistical-mechanical descriptions of large populations, although the microscopic construction and the role of inter-player interactions depend strongly on the model considered.

At the microscopic level, the \textit{Eisert–Wilkens–Lewenstein} (EWL) protocol \cite{ref15} gives us a natural quantum-game building block. An initially separable two-qubit state is entangled, acted on by local strategic unitaries, disentangled, and measured. The protocol and its strategy interpretation have subsequently been reviewed and
generalized~\cite{ref16}. Entanglement can modify payoff
orderings, Nash equilibria, and evolutionary stability, although such conclusions can depend on the allowed quantum strategy space~\cite{ref17}. The strategy set (details in Sec.~\ref{theory}) considered here is therefore specified explicitly rather than interpreting the results as statements about arbitrary $SU(2)$ strategies.

Physical decoherence introduces a second and independent source of \textit{stochasticity}. Quantum games subject to dephasing, depolarizing, amplitude-damping, and related channels have been studied since the early development of noisy quantum-game protocols~\cite{ref18, ref19}. Later studies considered several circuit conventions, including noisy evolution in stages surrounding the players' strategic operations, and examined channel dependence, memory effects, payoff surfaces, and Nash equilibria~\cite{ref20, ref21, ref22}. Noise has also been studied in finite multiplayer quantum games, including general multiplayer
constructions and noisy three-player dilemmas~\cite{ref23, ref24}. More recent \textit{EWL}-based multiplayer work has derived symmetric Nash equilibria and entanglement thresholds in the quantum Volunteer's dilemma for finite $n$~\cite{ref44, ref45}. In a different direction, Hamiltonian-driven quantum game constructions based on transverse-field Ising dynamics have used a modified \textit{EWL} scheme to generate tunable entanglers and to analyze the strategic equilibria, including the Stag Hunt, with subsequent work treating amplitude damping decoherence~\cite{ref46, ref47}. These results show that neither decoherence nor finite multiplayer quantum games are new ingredients by themselves. The goal of this paper is more specific: a \emph{single} channel instance is applied and its position is moved from the \textit{resource} segment to the \textit{post-strategy} segment. The resulting noisy two-player payoff matrix is then used as the local interaction of a short-range equilibrium population with $N=L^2$ strategic degrees of freedom.

The \textit{Stag Hunt} is particularly suitable for this purpose because it is a coordination game with competing cooperative and safe equilibria~\cite{ref25}. Quantum Stag Hunt games have been studied at the two-player level~\cite{ref26, ref27}, while quantum strategies have also been embedded in regular lattices and evolving networks~\cite{ref28, ref29}. Spatial noisy quantum games have also been considered in cellular automaton and evolutionary settings~\cite{ref30}. A recent graph-encoded multiplayer construction is especially relevant geometrically: Tsakiroglou \emph{et al.} associate two qubits with each graph edge and treat each edge as a distinct two-player quantum game~\cite{ref31}. Their setting, however, uses ideal state-vector simulations together with adaptive EXP3 strategy updates, while decoherence and explicit noise-model simulations are left as future extensions. It therefore does not reduce a physically noisy payoff matrix to an equilibrium Ising pair potential or study short-range thermodynamic criticality. These earlier constructions overlap with individual ingredients of the present model, but not with the complete channel-to-interaction-to-criticality sequence considered below.

A complementary direction starts from potential games and develops statistical mechanical descriptions of phase transition-like behavior, including quantum-game examples in highly connected populations~\cite{ref32}. This also emphasizes an important distinction: a change in Nash-equilibrium structure and a thermodynamic phase transition are not the same object. That distinction is maintained here. The critical point considered below is defined by a singularity of the short range two-dimensional (\textit{2D}) Gibbs system, while the underlying two-player game can retain the same set of pure Nash equilibria across the microscopic neutrality point.

The question addressed in this work is therefore deliberately narrow: how does a physical channel acting inside the \textit{EWL} circuit renormalize the effective pair potential of a short range equilibrium
population, and can different channels, or different placements of the same channel, place that population in different thermodynamic regimes? 

The construction follows this sequence:
\begin{equation}
  \varepsilon_p
  \to \bar{\Delta}_{\varepsilon}(\Gamma,p)
  \to  \big[\mathfrak{J}_{\varepsilon}(\Gamma,p), \mathfrak{H}_{\varepsilon}(\Gamma,p)\big] \to P(\{\mathfrak{s}_i\}) \propto e^{-\beta\mathsf{H}} .
  \label{eq1}
\end{equation}
The parameter $p$ quantifies microscopic physical decoherence, whereas $\beta^{-1}$ sets the scale of classical strategic, or \textit{Gibbs}, fluctuations. Keeping these two sources of randomness separate is essential for isolating the effect of the physical channel.

The choice of payoff matrix is also deliberate. The reduced $\mathsf{Q/D}$ donation-game parametrization used in Ref.~\cite{ref14} lies on the special surface $\mathsf{R+P=S+T}$ (these will be discussed in Sec.~\ref{theory}), for which the interaction obtained from the payoff differences satisfies $\mathfrak{J}=0$. The Stag Hunt matrix considered here instead gives $\mathfrak{J}=3/4$ in the noiseless limit and therefore supports collective nearest neighbor (N.N.) ordering. Physical decoherence can consequently modify a nonzero interaction rather than only an effective field.

The main results are as follows: closed form expressions for $\mathfrak{J}(\Gamma,p)$ and $\mathfrak{H}(\Gamma,p)$ are obtained for
phase damping, depolarization, and amplitude damping at two single channel circuit placements, and are independently verified using a direct
density matrix/Kraus calculation. Phase damping and depolarization share the same microscopic neutrality curve $\mathfrak{H}=0$, but only depolarization suppresses the interaction as $(1-p)^2$. At fixed $\beta=1$, the depolarizing branch therefore reaches the exact square lattice critical point $p = p_{*}\approx 0.233460\ldots$, where the field driven coexistence line terminates. Zero-field Monte Carlo simulations provide an independent check of this analytical prediction, while nonzero-field $\Gamma$ scans distinguish two-phase coexistence for $p<p_{*}$ from a smooth crossover for $p>p_{*}$. Amplitude damping gives a separate placement-dependent result: moving the channel across the strategic layer changes the effective game, and the post-strategy neutrality branch reaches $\Gamma=0$ at $p=1/3$ before losing its nontrivial interacting solution. Throughout, the extended lattice remains a \textit{classical} Ising system; quantum mechanics enters through the channel dependent interaction generated by each microscopic \textit{EWL} encounter.

\section{\label{theory}Theory}
\subsection{\label{subsec2a}Classical Stag-Hunt \& its equilibrium structure}
We start with the classical symmetric Stag-Hunt payoff matrix,
\begin{equation}
    \Delta_{S-H} = \begin{bmatrix}
    \mathsf{R} & \mathsf{S}\\
    \mathsf{T} & \mathsf{P}
    \end{bmatrix} 
    = \begin{bmatrix}
    4 & 0\\
    3 & 2
    \end{bmatrix},~\text{where}~\{\mathsf{R}>\mathsf{T} > \mathsf{P} > \mathsf{S}\}.
    \label{eq2}
\end{equation}
In this case, the \textit{first} and \textit{second} actions may be considered as \textit{stag/cooperation} $(\mathsf{C})$ and \textit{hare/defection} $(\mathsf{D})$, respectively. This game has two well-known pure coordination equilibria: $(\mathsf{C,C})$ and $(\mathsf{D,D})$ \cite{ref25}. We can notice that if the opponent cooperates with probability $\mathfrak{x}$, then the expected \textit{cooperation} payoff is $4\mathfrak{x}$ whereas the \textit{defection} payoff is $3\mathfrak{x}+2(1-\mathfrak{x}) = \mathfrak{x} + 2$, indicating that the interior \textit{mixed} equilibrium happens at,
\begin{equation}
    \mathfrak{x}_{*} = \dfrac{\mathsf{P-S}}{\mathsf{R+P-S-T}} = \dfrac{2}{3}.
    \label{eq3}
\end{equation}
For this payoff choice, we observe that mutual \textit{cooperation} gives the better outcome, but the \textit{defection} (hare) equilibrium is easier to achieve and is therefore \textit{risk dominant}. These classical features are very useful for understanding the game, but the \textit{thermodynamic} transition that will be discussed next is based on the effective potential of the noisy and quantized \textit{quantum}($\mathsf{Q}$)/\textit{defection}($\mathsf{D}$) game, rather than by the classical ``\textit{mixed}" equilibrium.

\subsection{\label{subsec2b}EWL quantization \& the $\mathsf{Q/D}$ sector}
The \textit{EWL} protocol links the two \textit{classical} actions with the computational basis states: $|\mathsf{C}\rangle \equiv |0\rangle$ and $|\mathsf{D}\rangle \equiv |1\rangle$, and allows each of the two players (say, \textit{A} and \textit{B}) to act locally on a \textit{two}-qubit resource \cite{ref15, ref16}. For the generic \textit{two}-parameter strategy case,
\begin{gather}
    \mathbb{U(\alpha, \delta)} = \begin{bmatrix}
        \cos{(\frac{\alpha}{2})} \exp{(\mathsf{i}\delta)} & \sin{(\frac{\alpha}{2})}\\ 
        -\sin{(\frac{\alpha}{2})} & \cos{(\frac{\alpha}{2})} \exp{(-\mathsf{i}\delta)}
    \end{bmatrix},
    \nonumber\\
    \forall~ \alpha \in [0, \pi]~~\text{and}~~\delta \in \bigg[0, \frac{\pi}{2}\bigg],
    \label{eq4}
\end{gather}
such that, 
\begin{equation}
    \mathsf{C} = \mathbb{U}(0,0) = \mathbb{I}, ~\mathsf{D} = \mathbb{U}(\pi,0) = i\mathbb{Y},~\mathsf{Q} = \mathbb{U}\bigg(0,\frac{\pi}{2}\bigg) = i\mathbb{Z},
    \label{eq5}
\end{equation}
where $\{{\mathbb{I,Y,Z}}\}$ are the regular $2\times2$ \textit{Pauli} matrices. Here, we deliberately restrict our game to the $\{\mathsf{Q,D}\}$-subspace so that we can establish a clean mapping between the two available strategies and the Ising \textit{spin} variables, as discussed in the following sections. This allows the resulting $2\times2$ payoff matrix to be mapped onto the effective Ising parameters $\mathfrak{J}$ and $\mathfrak{H}$ (see Refs.~\cite{ref12,ref14} for more details). Such restricted \textit{EWL} strategy sets have been previously used in the quantum-game literature, but they should not be viewed as the full $SU(2)$-strategy space. Recent studies have shown explicit cases in which the strategic stability can change drastically upon the enlargement of the allowed strategy space \cite{ref17}. 

In the \textit{EWL} protocol, the \textit{entangler} is given as \cite{ref15},
\begin{equation}
    \mathcal{J} (\Gamma) = \cos{\frac{\Gamma}{2}} ~(\mathbb{I \otimes I}) - \mathsf{i} \sin{\frac{\Gamma}{2}} ~(\mathbb{Y \otimes Y}),~\forall~\Gamma \in \bigg[0,\frac{\pi}{2}\bigg],
    \label{eq6}
\end{equation}
and this gives us the \textit{entangled} state,
\begin{equation}
    |\Omega_1\rangle = \mathcal{J} (\Gamma)  |00\rangle\equiv \mathcal{J} (\Gamma)  |\mathsf{CC}\rangle = \cos{\frac{\Gamma}{2}} |00\rangle + \mathsf{i} \sin{\frac{\Gamma}{2}} |11\rangle,
    \label{eq7}
\end{equation}
where we assumed that the \textit{initial state} of both the players were $|0\rangle$ (or, $|\mathsf{C}\rangle$). Here, the endpoint values of $\Gamma =0$ and $\frac{\pi}{2}$ signify a \textit{separable} and \textit{maximally entangled} resource, respectively. For a given final density matrix $\varrho_f$, the payoff operators are \cite{ref15},
\begin{gather}
    \hat{\Pi}_A = \mathsf{R} |00\rangle \langle00| + \mathsf{S} |01\rangle \langle01| + \mathsf{T} |10\rangle \langle10| + \mathsf{P} |11\rangle \langle11| \label{eq8},\\
    \hat{\Pi}_B = \mathsf{R} |00\rangle \langle00| + \mathsf{T} |01\rangle \langle01| + \mathsf{S} |10\rangle \langle10| + \mathsf{P} |11\rangle \langle11| \label{eq9},
\end{gather}
and we have the expected payoffs as,
\begin{equation}
    \Pi_A = \text{Tr}(\varrho_f~\hat{\Pi}_A ),~~\Pi_B = \text{Tr}(\varrho_f~\hat{\Pi}_B ).
    \label{eq10}
\end{equation}
Equivalently, for the \textit{row} player,
\begin{equation}
    \Pi_A = \mathsf{R} \mathbb{P}_{00}  + \mathsf{S} \mathbb{P}_{01}+ \mathsf{T} \mathbb{P}_{10} + \mathsf{P} \mathbb{P}_{11},
    \label{eq11}
\end{equation}
where $\mathbb{P}_{mn} = \langle mn| \varrho_f |mn\rangle$. Without physical noise, the final state is given as \cite{ref15},
\begin{equation}
    \varrho_f = \mathcal{J}^\dagger (\Gamma) (\mathbb{U}_A \otimes \mathbb{U}_B) |\Omega_1\rangle\langle \Omega_1| (\mathbb{U}_A \otimes \mathbb{U}_B)^\dagger \mathcal{J} (\Gamma)
    \label{eq12}
\end{equation}
For a general symmetric payoff matrix ${(\mathsf{R,S,T,P})}$, evaluating the four $\mathsf{Q/D}$-strategy pairs gives us,
\begin{equation}
    \bar{\Delta}_{\mathsf{QD}}(\Gamma) = \begin{bmatrix}
        \mathsf{R} & \mathsf{S}\cos^2(\Gamma) + \mathsf{T}\sin^2(\Gamma)\\
        \mathsf{T}\cos^2(\Gamma) + \mathsf{S}\sin^2(\Gamma) & \mathsf{P}
    \end{bmatrix} 
    \label{eq13}
\end{equation}
For Eq.~(\ref{eq2}),
\begin{equation}
    \bar{\Delta}^{(0)}_{\mathsf{QD}}(\Gamma) = \begin{bmatrix}
        4 & 3\sin^2(\Gamma)\\
        3\cos^2(\Gamma) & 2
    \end{bmatrix} 
    \label{eq14}
\end{equation}
At $\Gamma=0$, $\mathsf{Q}$ is payoff-equivalent to the classical \textit{cooperative} action. Hence, for our case, Eq.~(\ref{eq14}) reduces to Eq.~(\ref{eq2}). A detailed calculation is included in Appendix~\ref{appendixA}

\subsection{\label{subsec2c}Open-system channel \& circuit placement}
For our case, physical \textit{decoherence} is introduced as a completely positive trace-preserving (CPTP) map in Kraus form \cite{ref33},
\begin{equation}
    \varepsilon_{p}(\varrho) = \sum_{a} \mathsf{K}_a \varrho \mathsf{K}_a^\dagger, ~~\sum_a \mathsf{K}_a^\dagger \mathsf{K}_a = \mathbb{I},
    \label{eq15}
\end{equation}
with \textit{independent, identical} channels on the two \textit{EWL} qubits,
\begin{equation}
    \varepsilon_{p} ^{\otimes 2} (\varrho) = \sum_{a,b} (\mathsf{K}_a \otimes \mathsf{K}_b) \varrho (\mathsf{K}_a \otimes \mathsf{K}_b)^\dagger.
    \label{eq16}
\end{equation}
In Eq.~(\ref{eq15}), $p \in [0,1]$ denotes the \textit{strength} of the physical decoherence channel, with $p = 0$ indicating the noiseless limit. Its precise meaning depends on the type of channel that we take into consideration \cite{ref33}.

Phase damping, depolarization, and amplitude damping are standard prototype channels in open-system quantum information and in noisy quantum games \cite{ref19, ref20, ref22, ref23}. We use the following convention so that the numerical value of $p$ is unambiguous.

For phase damping,
\begin{equation}
    \mathsf{K}_0^{ph} = \begin{pmatrix}
        1 & 0\\
        0 & \sqrt{1-p}
    \end{pmatrix},~~~
    \mathsf{K}_1^{ph} = \begin{pmatrix}
        0 & 0\\
        0 & \sqrt{p}
    \end{pmatrix}.
    \label{eq17}
\end{equation}
This channel preserves the computational basis populations while suppressing the one-qubit coherence by a factor of $\sqrt{1-p}$. For depolarization, our convention is,
\begin{equation}
    \mathcal{D}_p(\varrho) = (1-p)\varrho + p\frac{\mathbb{I}}{2},
    \label{eq18}
\end{equation}
which is implemented via,
\begin{equation}
    \mathsf{K}_0^{dep} = \sqrt{1-\frac{3p}{4}}~\mathbb{I},~~\mathsf{K}_j^{dep} = \sqrt{\frac{p}{4}}~{\varsigma_j},~~\varsigma_j\in \{{\mathbb{X,Y,Z}}\} 
    \label{eq19}
\end{equation}
Previous noisy-game literature dealt with different normalizations of the depolarizing parameter, so numerical thresholds should only be compared after the channel conventions are matched \cite{ref20}. 

For amplitude damping,
\begin{equation}
    \mathsf{K}_0^{ad} = \begin{pmatrix}
        1 & 0\\
        0 & \sqrt{1-p}
    \end{pmatrix},~~~
    \mathsf{K}_1^{ad} = \begin{pmatrix}
        0 & \sqrt{p}\\
        0 & 0
    \end{pmatrix},
    \label{eq20}
\end{equation}
which is non-unital and relaxes $|1\rangle$ toward $|0\rangle$. 

\begin{figure}
    \centering
    \includegraphics[width=1\linewidth]{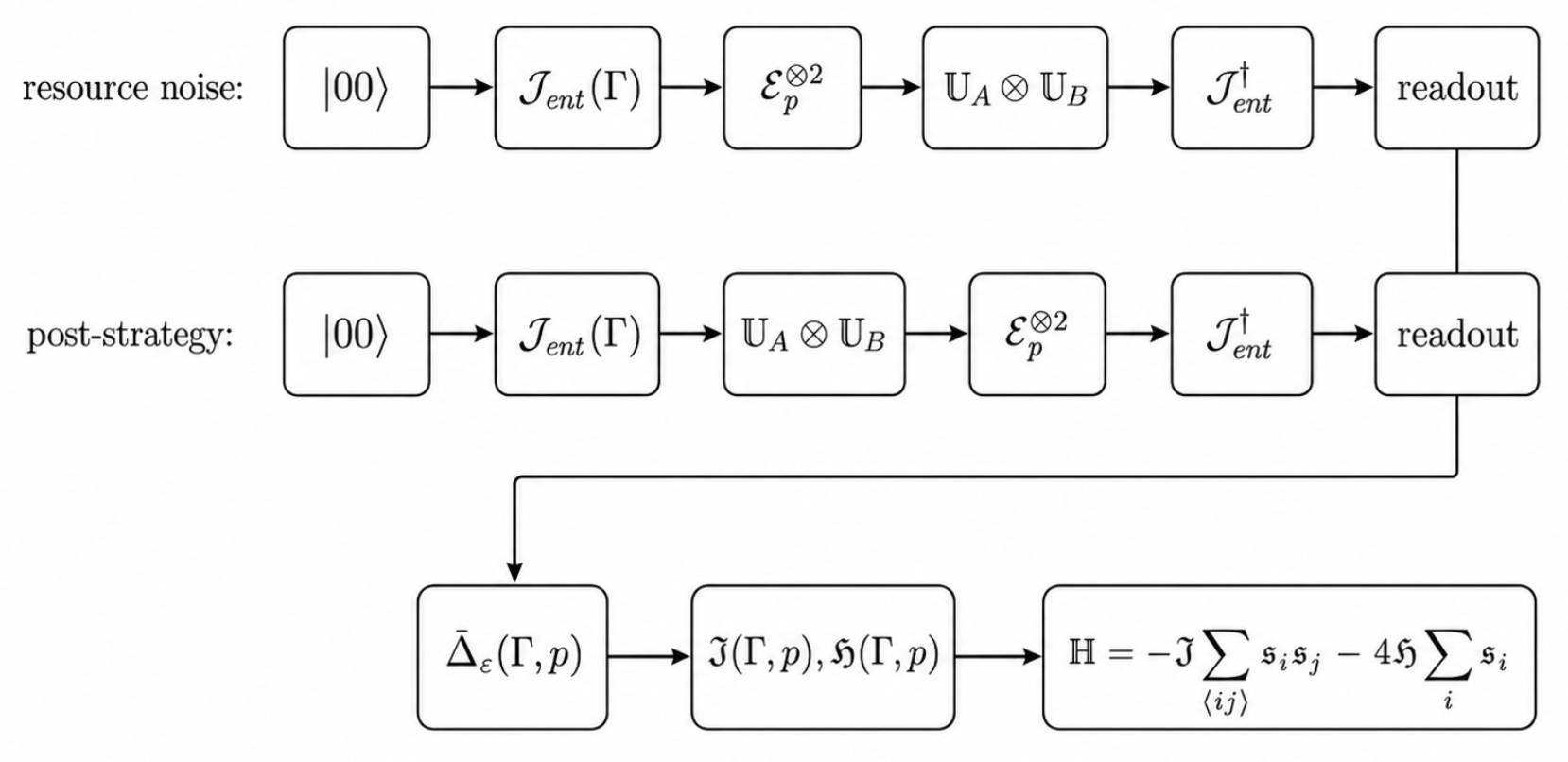}
    \caption{Microscopic-to-collective flowchart: each N.N. bond of the square lattice represents an independent noisy \textit{EWL} encounter. A site carries one strategic variable $\mathfrak{s}_i=+1$ ($\mathsf{Q}$) or $-1$ ($\mathsf{D}$), shared across its \textit{four} bonds. The quantum channel is applied either to the entangled resource or after the local strategic operations. The resulting $2\times2$ payoff matrix determines the classical pair interaction $\mathfrak{J}$ and field $\mathfrak{H}$.}
    \label{fig1}
\end{figure}

We compare two different cases of single noise channel insertions. Prior noisy-game protocols considered decoherence in circuit segments both \textit{before} and \textit{after} the strategic operations, including noisy evolution in more than one segment \cite{ref20, ref22}. However, our approach is slightly different -- we isolate the position of one channel application such that moving it across the $\mathsf{Q/D}$-strategy layer is the only change. In the \emph{resource-noise} ordering,
\begin{gather}
  \varrho_1 = \mathcal{J}(\Gamma)\varrho_0 \mathcal{J}^\dagger (\Gamma),~~
  \varrho_2 =\varepsilon_p^{\otimes2}(\varrho_1),\nonumber\\
  \varrho_3 =(\mathbb{U}_A\otimes \mathbb{U}_B)\varrho_2(\mathbb{U}_A\otimes \mathbb{U}_B)^\dagger,~~
  \varrho_f = \mathcal{J}^\dagger (\Gamma)\varrho_3 \mathcal{J} (\Gamma),
  \label{eq21}
\end{gather}
where $\varrho_0=|00\rangle\langle 00|$. Meanwhile, in the \emph{post-strategy} ordering,
\begin{gather}
  \varrho_1 = \mathcal{J}(\Gamma)\varrho_0 \mathcal{J}^\dagger (\Gamma),~~
  \varrho_2 =(\mathbb{U}_A\otimes \mathbb{U}_B)\varrho_1(\mathbb{U}_A\otimes \mathbb{U}_B)^\dagger,\nonumber\\
  \varrho_3 = \varepsilon_p^{\otimes2}(\varrho_2),~~
  \varrho_f = \mathcal{J}^\dagger (\Gamma)\varrho_3 \mathcal{J} (\Gamma).
  \label{eq22}
\end{gather}
The two distinct paths and their connection to the lattice model are summarized in Fig.~\ref{fig1}. This single-insertion approach is intentionally distinct from double-segment noisy-game protocols in which decoherence acts in more than one circuit stage \cite{ref20}.

%% ------ Fig. 1 here ------

\subsection{\label{subsec2d}Exact representation of the symmetric game}
Let the \textit{row} player payoff matrix after the  quantum circuit be,
\begin{equation}
    \bar{\Delta} = \begin{bmatrix}
        \mathit{m} & \mathit{n}\\
        \mathit{r} & \mathit{q}
    \end{bmatrix}.
    \label{eq23}
\end{equation}
For a symmetric two-player game, the \textit{column} player's matrix is simply $\bar{\Delta}^{T}$. Such a binary symmetric game is defined as an exact potential game: there exists a scalar pair potential whose change under a unilateral strategy switch equates to that player's payoff change \cite{ref1}. Here, we identify $\mathsf{Q}$ with $\mathfrak{s}_i =+1$ and $\mathsf{D}$ with $\mathfrak{s}_i = -1$, and choose,
\begin{equation}
    \Phi_{ij} = \mathfrak{J}\mathfrak{s}_i\mathfrak{s}_j + \mathfrak{H}(\mathfrak{s}_i+\mathfrak{s}_j) + \Phi_0,
    \label{eq24}
\end{equation}
where the irrelevant constant $\Phi_0$ may be set to zero. Hence, we have the corresponding potential matrix as,
\begin{equation}
    \Phi = \begin{bmatrix}
        \mathfrak{J+2H} & \mathfrak{-J}\\
        \mathfrak{-J} & \mathfrak{J-2H}
    \end{bmatrix}.
    \label{eq25}
\end{equation}
Exact-potential consistency requires,
\begin{gather}
    \Phi_{\mathsf{QQ}} - \Phi_{DQ} = \mathit{m-r},~~
    \Phi_{\mathsf{QD}} - \Phi_{DD} = \mathit{n-q},
    \label{eq26}
\end{gather}
and this gives us,
\begin{equation}
    \mathfrak{J} = \frac{\mathit{m}+\mathit{q}- \mathit{r} - \mathit{n}}{4},~~\mathfrak{H} = \frac{\mathit{m}+\mathit{n}- \mathit{r} - \mathit{q}}{4}.
    \label{eq27}
\end{equation}
Hence, we have $2(\mathfrak{J+H}) = \mathit{m}-\mathit{r}$ and $2(\mathfrak{J-H}) = \mathit{q}-\mathit{n}$. Eq.~(\ref{eq27}) is algebraically similar to the payoff-difference technique employed in NEM and related game-to-Ising mappings \cite{ref5, ref11, ref12}, but expressing it as an exact bond potential makes the multiple-neighbor extension more intuitive. We can also notice that only unilateral payoff differences matter; adding a constant to all the payoffs in a column has no effect on the strategic potential. 

For the general noiseless \textit{EWL} payoff matrix in Eq.~(\ref{eq13}), Eq.~(\ref{eq27}) gives us,
\begin{equation}
    \mathfrak{J}^{(0)} = \frac{\mathsf{R+P-S-T}}{4}, ~\mathfrak{H}^{(0)}(\Gamma) = \frac{\mathsf{R-P} + (\mathsf{S-T})\cos{2\Gamma}}{4}
    \label{eq28}
\end{equation}
For our choice of parameter values in Eq.~(\ref{eq2}), the Stag-Hunt game gives us,
\begin{equation}
    \mathfrak{J}^{(0)} = \frac{3}{4}, ~\mathfrak{H}^{(0)}(\Gamma) = \frac{2 -3\cos{2\Gamma}}{4}
    \label{eq29}
\end{equation}
We might notice in Eq.~(\ref{eq28}), for $\mathsf{R+P>S+T}$, we have $\mathfrak{J}>0$ and this leads to a \textit{ferromagnetic} coordination tendency, whereas for $\mathsf{R+P<S+T}$, we have $\mathfrak{J}<0$ and this instead leads to an \textit{antiferromagnetic} or anti-coordination tendency \cite{ref5}. The current game lies in the former class and this distinguishes itself from other games that satisfy either $\mathfrak{J}< 0$ or $\mathfrak{J} =0$ (see, Ref.~\cite{ref14}).

\subsection{\label{subsec2e}From pair potential to \textit{2D} Gibbs population}
Following the previous discussion, we now place one strategic variable $\mathfrak{s}_i = \pm 1$ on each site of a regular graph and relate each undirected N.N. edge $\langle ij \rangle$ with an independent \textit{EWL} protocol having the same $(\Gamma, p)$ and channel convention. The player's strategy is considered the same in all of its bond encounters, while the resources and channel realizations are edge-independent. The total payoff for a player is determined by summing over all the nearest neighbors. Assigning distinct two-qubit game resources to graph edges has been studied extensively in graph-encoded multiplayer quantum game circuits \cite{ref31}; the current discussion differs by considering a common site strategy across all the incident bonds and by relating the \textit{noisy} edge payoff matrix to an equilibrium potential. When we say that the edge resources are \textit{independent}, we do \textit{not} mean that a single physical qubit at the $i^{th}$-site is simultaneously entangled with all of its neighbors. A \textit{simultaneous} realization on the $2D$ square lattice requires us to consider four local qubit subsystems per player, one for each bond, while an equivalent \textit{sequential} realization may repeatedly prepare independent two-qubit resources. In either case, the variable $\mathfrak{s}_i$ records the common strategic choice applied to the four bond encounters. 

Under these assumptions, the total (\textit{global}) game potential is the sum of the individual game potentials,
\begin{equation}
    \Phi_{total} = \sum_{\langle ij \rangle}[\mathfrak{J}\mathfrak{s}_i\mathfrak{s}_j + \mathfrak{H}(\mathfrak{s}_i+\mathfrak{s}_j)].
    \label{eq30}
\end{equation}
For a regular graph of coordination number $z$, we have,
\begin{equation}
    \sum_{\langle ij\rangle} (\mathfrak{s}_i+\mathfrak{s}_j) = z\sum_{i} \mathfrak{s}_i,
    \label{eq31}
\end{equation}
so the Gibbs Hamiltonian $\mathsf{H} = -\Phi_{total}$ is,
\begin{equation}
    \mathsf{H} = -\mathfrak{J}\sum_{\langle ij \rangle} \mathfrak{s}_i\mathfrak{s}_j - z\mathfrak{H}\sum_{i} \mathfrak{s}_i.
    \label{eq32}
\end{equation}
For a square lattice, we have $z = 4$. Hence, 
\begin{equation}
    \mathsf{H} = -\mathfrak{J}\sum_{\langle ij \rangle} \mathfrak{s}_i\mathfrak{s}_j - 4\mathfrak{H}\sum_{i} \mathfrak{s}_i.
    \label{eq33}
\end{equation}
In Eq.~(\ref{eq33}), the factor $4\mathfrak{H}$ is \textit{not} an additional phenomenological field; instead it follows from summing the same pair potential over the four bonds incident on each site. Currently, each player's payoff is determined by summing over the payoffs obtained from its four N.N. encounters. Instead, if we considered the average payoff, then the entire game potential would be divided by four, which equivalently corresponds to a rescaling of the inverse strategic noise parameter $\beta$.

For an exact potential game with logit-type stochastic choice, the stationary distribution takes the Gibbs form in the game potential \cite{ref2, ref7}. We consider the equivalent equilibrium ensemble,
\begin{equation}
    P(\{ \mathfrak{s}_i \}) = \frac{1}{\mathcal{Z}} \exp{[-\beta \mathsf{H} (\{ \mathfrak{s}_i  \})]}, ~~\mathcal{Z} = \sum_{\{ \mathfrak{s}_i \}} \exp{[-\beta \mathsf{H}]}.
    \label{eq34}
\end{equation}
Here, $\beta$ can be defined as an inverse strategic noise parameter, as in earlier thermodynamic limit game models \cite{ref8,ref10,ref11,ref14}. Larger $\beta$ suppresses strategic randomness, while the limit $\beta \rightarrow 0$ makes the Gibbs weights increasingly insensitive to payoff differences. It is independent, both conceptually and operationally, of the quantum channel parameter $p$. The CPTP map is evaluated at the density matrix level before the lattice model is constructed, so the Monte Carlo algorithm used later is a sampler of Eq.~(\ref{eq34}); it should not be interpreted as a universal microscopic model of strategic behavior. It does not sample individual Kraus trajectories or microscopic channel realizations.

\subsection{\label{subsec2f}Nash equilibria against potential neutrality}
The readers might get confused with the language of \textit{phase transitions} and the two-player equilibrium changes. Hence, it is useful to separate the two notions explicitly \cite{ref32}. For a general symmetric payoff matrix in Eq.~\eqref{eq23}, Eq.~\eqref{eq27} gives us,
\begin{gather}
    (\mathsf{Q,Q}) ~\text{is a Nash equilibrium}~\iff \mathfrak{J+H \geq} ~0,\nonumber \\
    (\mathsf{D,D}) ~\text{is a Nash equilibrium}~\iff \mathfrak{J-H \geq}~0.
\end{gather}
Thus, for a coordination game with $\mathfrak{J} > 0$, both pure coordination equilibria coexist whenever $|\mathfrak{H}| \leq \mathfrak{J}$. When an interior mixed equilibrium exists, equality of the two expected payoffs gives us,
\begin{equation}
    x^{*} = \frac{1}{2} - \frac{\mathfrak{H}}{2\mathfrak{J}},
\end{equation}
where $x^{*}$ is the probability that opponent chooses $\mathsf{Q}$. Hence $\mathfrak{H} = 0$ is the equal potential (risk-dominance-neutral) point $x^{*} = \frac{1}{2}$; it is not in general a Nash equilibrium bifurcation. For the noiseless matrix in Eq.~(\ref{eq14}), the $(\mathsf{Q,Q})$ profile becomes the Nash equilibrium when,
\begin{equation}
    3 \cos^2{\Gamma} \leq 4,
    \label{eq35}
\end{equation}
which is satisfied for all $\Gamma \in [ 0, \frac{\pi}{2}]$. The $(\mathsf{D,D})$ profile is also a Nash equilibrium given that,
\begin{equation}
    3 \sin^2{\Gamma} \leq 2 \iff \sin^2{\Gamma} \leq \frac{2}{3}.
    \label{eq36}
\end{equation}
We also notice that the potential \textit{neutrality} condition is $\mathfrak{H}^{(0)}(\Gamma) = 0$ (see, Eq.~(\ref{eq29})), or,
\begin{equation}
    \sin^2{\Gamma_c} = \frac{1}{6},
    \label{eq37}
\end{equation}
and this lies well within the region where both of the pure coordination equilibria coexist. Hence, the field-driven coexistence line, that we will study further, does not equate to a loss of stability of the two-player Nash equilibria. Rather, it marks the \textit{``point"} where the two homogeneous potential phases are thermodynamically \textit{degenerate} and they carry equal Gibbs weightage. This distinction also remains important once physical decoherence is introduced.

\section{\label{sec3}Exact Channel-dependent interaction}
\begin{table*}[t]
\renewcommand{\arraystretch}{1.85}
\resizebox{\textwidth}{!}{%
\begin{tabular}{lll}
\toprule
\textbf{channel}/\textbf{\textit{placement}} & $\mathfrak{J}(\Gamma,p)$ & $\mathfrak{H}(\Gamma,p)$\\
\midrule
phase damping, \textit{either}
& $\dfrac{3}{4}$
& $\dfrac{2-3\cos{2\Gamma} - 5p\sin^2\Gamma}{4}$\\[2pt]

depolarizing, \textit{either}
& $ \dfrac{3}{4}(1-p)^2$
& $\dfrac{1-p}{4}\left[2-3\cos{2\Gamma}-5p\sin^2\Gamma\right]$\\[2pt]

amplitude damping, \textit{resource}
& $\dfrac34\left[1-2p(1-p)(1-\cos\Gamma)\right]$
& $\dfrac{2-3\cos{2\Gamma}-p(1-\cos\Gamma)(5+6\cos\Gamma)}{4}$\\[2pt]

amplitude damping, \textit{post-strategy}
& $ \dfrac{3}{4}(1-p)^2$
& $ \dfrac{1-p}{4}\left[2-3\cos2\Gamma+3p\cos\Gamma\right]$\\
\bottomrule
\end{tabular}%
}
\caption{Exact effective Ising parameters for the Stag-Hunt matrix in Eq.~\eqref{eq2}. $(\mathcal{R})$ and $(\mathcal{S})$ indicate \textit{resource} and \textit{post-strategy} amplitude damping, respectively (see the main text).}
\label{table1}
\end{table*}

\subsection{\label{subsec3a}Closed-form Ising structure}
The full noisy payoff matrices are determined by considering Eqs.~(\ref{eq21}) and (\ref{eq22}) for all four $\mathsf{Q/D}$ pairs and then applying Eq.~(\ref{eq27}). The main closed-form result is summarized in Table~\ref{table1}; full payoff matrices are included in Appendix~\ref{appendixB}.

Two structural facts are easily noticed. First, phase damping changes $\mathfrak{H}$ but leaves the coupling/interaction strength at exactly $\mathfrak{J} = \frac{3}{4}$. Depolarization produces a similar effect in $\mathfrak{H}$ (see Table~\ref{table1}) but also suppresses $\mathfrak{J}$ by a factor of $(1-p)^2$. For the restricted construction and the convention used here, this can also be written as $\mathfrak{J}_{\text{dep}}(p) = (1-p)^2 \mathfrak{J}^{(0)}$. Second, amplitude damping depends on where it acts in the \textit{EWL} circuit.

The placement equivalence of phase damping and depolarization is a \textit{covariance} property of the restricted strategy set that we take into consideration, rather than a numerical coincidence. A channel $\varepsilon$ is considered \textit{covariant} with respect to a unitary $\mathbb{U}$ when $\varepsilon(\mathbb{U}\varrho \mathbb{U}^\dagger) = \mathbb{U}\varepsilon(\varrho)\mathbb{U}^\dagger$ \cite{ref33}. The depolarizing channel is unitarily covariant, i.e.,
\begin{equation}
    \mathcal{D}_p(\mathbb{U}\varrho \mathbb{U}^\dagger) = \mathbb{U}\mathcal{D}_p(\varrho)\mathbb{U}^\dagger,
    \label{eq38}
\end{equation}
so $\mathcal{D}_p^{\otimes 2}$ can be commuted through $\mathbb{U}_A \otimes \mathbb{U}_B$ for any local strategies. The chosen phase-damping channel is not fully unitarily covariant, but it is indeed covariant under the Pauli operations $\mathsf{Q} \propto \mathbb{Z}$ and $\mathsf{D} \propto \mathbb{Y}$; hence the commutation holds on the restricted $\mathsf{Q/D}$ sector. Generally, amplitude damping is non-unital and not Pauli covariant since,
\begin{equation}
    \varepsilon_{ad} (\mathbb{Y}\varrho \mathbb{Y} ) \neq \mathbb{Y}\varepsilon_{ad}(\varrho) \mathbb{Y},~~\mathbb{Y} = \mathbb{Y}^\dagger.
    \label{eq39}
\end{equation}
Hence, we observe that resource and post-strategy amplitude damping generate different effective games. Earlier studies have established that location and sequencing of decoherence in an \textit{EWL}-type circuit can affect finite-player payoffs \cite{ref20, ref22}, whereas, our study, where we insert the noise at a single location, isolates that effect at the level of the effective pair potential.

\subsection{\label{subsec3b}Exact \textit{zero}-field lines}
The line $\mathfrak{H}(\Gamma, p) = 0 $ indicates the microscopic strategic neutrality, i.e., neither of the two magnetized strategic orientations is favored by the effective field. For the noiseless case,
\begin{equation}
    \Gamma_c (0) = \frac{1}{2} \cos^{-1}{\frac{2}{3}} \approx 0.420534\ldots.
    \label{eq40}
\end{equation}
For both phase damping and depolarizing noise channels,
\begin{equation}
    \sin^{2}{\Gamma_c}(p) = \frac{1}{6 - 5p}, ~~\cos{2\Gamma_c}(p) = \frac{4-5p}{6-5p}.
    \label{eq41}
\end{equation}
When $p=1$, we have $\mathfrak{J}=\mathfrak{H}=0$ for all $\Gamma$ in the depolarizing channel. Thus, the endpoint of Eq.~\eqref{eq41} is \emph{not} an interacting critical point; rather, $\Gamma=\pi/2$ is the continuous endpoint of the interacting neutrality branch as $p\to1^{-}$.

For \textit{resource} amplitude damping, we define $\mathfrak{c}_{\mathcal{R}} = \cos{\Gamma_c}^{(\mathcal{R})}$ and the zero-field equation is,
\begin{equation}
    (1-p)(5-6 \mathfrak{c}_{\mathcal{R}}^2) - p\mathfrak{c}_{\mathcal{R}} = 0,
    \label{eq42}
\end{equation}
with the physical roots being at,
\begin{equation}
    \mathfrak{c}_{\mathcal{R}}(p) = \frac{-p + \sqrt{p^2 + 120(1-p)^2}}{12(1-p)}, ~~ p \in [0,1),
    \label{eq43}
\end{equation}
and $\Gamma_c^{(\mathcal{R})}(p=1) = \frac{\pi}{2}$. The branch is valid for $p \in [0,1]$, since it extends continuously to $p=1$.

For \textit{post-strategy} amplitude damping, we define $\mathfrak{c}_{\mathcal{S}} = \cos{\Gamma_c}^{(\mathcal{S})}$ and the zero-field equation is,
\begin{equation}
    5-6 \mathfrak{c}_{\mathcal{S}}^2 + 3 p\mathfrak{c}_{\mathcal{S}} = 0,
    \label{eq44}
\end{equation}
with the physical roots being at,
\begin{equation}
    \mathfrak{c}_{\mathcal{S}}(p) = \frac{3p + \sqrt{9p^2 + 120}}{12} \leq 1 \implies p \leq \frac{1}{3}.
    \label{eq45}
\end{equation}
At $p=\frac{1}{3}$, the branch reaches $\Gamma_c^{\mathcal{(S)}} = 0$; for larger $p$ there is no interacting $\mathfrak{H} = 0$ point for $\Gamma \in [0, \frac{\pi}{2}]$.

\subsection{\label{subsec3c}Exact square-lattice criticality \& finite-size expectations}
When we have $\mathfrak{H} = 0$, Eq.~(\ref{eq33}) gives us the ferromagnetic square-lattice Ising model where the equilibrium behavior is controlled by the \textit{dimensionless} coupling, $\kappa = \beta \mathfrak{J}$. Larger values of $\kappa$ favor collective alignment of the Ising variables, while smaller values of $\kappa$ favor disorder. The thermodynamic critical coupling is given as \cite{ref34, ref35},
\begin{equation}
    \kappa_c = \frac{1}{2} \ln{(1+\sqrt{2})} \approx 0.440686\ldots .
    \label{eq46}
\end{equation}
Thus, $\kappa > \kappa_c$ is the ordered regime, $\kappa < \kappa_c$ is the disordered regime, and $\kappa = \kappa_c$ is the continuous \textit{2D} Ising critical point. 

For a periodic $L \times L$ lattice, $L$ denotes the system size along a linear direction and $N = L^2$ denotes the number of strategic variables. The signed magnetization per site is denoted by $\mu = N^{-1} \sum_{i} \mathfrak{s}_i$ and the finite size order parameter used below is $M = \langle |\mu|\rangle$. At zero field, the exact spontaneous magnetization is given by \cite{ref35, ref36},
\begin{equation}
  M_{L\rightarrow \infty}(\kappa)=
  \begin{cases}
  \left[1-\sinh^{-4}(2\kappa)\right]^{1/8},~\kappa >\kappa_c,\\[1mm]
  ~~~~~~~~~0,~~~~~~~~~~~\kappa \le \kappa_c.
  \end{cases}
  \label{eq47}
\end{equation}
We can use Eq.~(\ref{eq47}) as a useful analytical check for the Monte Carlo implementation even though it is not used in our main figures. 

To define the approach to criticality in a finite lattice, we use the reduced distance,
\begin{equation}
    t \equiv \frac{\kappa - \kappa_c}{\kappa_c},
    \label{eq48}
\end{equation}
such that $t=0$ at the critical point. When we are near the critical point, standard finite-size scaling for a periodic $L\times L $ system gives us \cite{ref37, ref38, ref39},
\begin{gather}
  M(t,L) = L^{-\beta_{\rm I}/\nu}\,\mathcal{M}(tL^{1/\nu})[1+\cdots]\label{eq49a},\\
  \chi_{|\mu|}(t,L) =L^{\gamma_{\rm I}/\nu}\,\mathcal{X}(tL^{1/\nu})[1+\cdots]\label{eq49b},\\
  U_4(t,L) =\mathcal{U}(tL^{1/\nu})[1+\cdots],
  \label{eq49c}
\end{gather}
where, $\chi_{|\mu|}$ is the susceptibility derived from the absolute magnetization, $U_4$ is the \textit{fourth}-order Binder cumulant, and $\mathcal{(M,X,U)}$ are finite-size scaling functions. Their explicit estimators are shown in Eqs.~(\ref{eq64}) and (\ref{eq65}). Also, $\beta_{\rm I},~\gamma_{\rm I}$, and $\nu$ are the magnetization, susceptibility and the correlation-length critical exponents, respectively. For the \textit{2D} Ising universality class,
\begin{equation}
    \beta_{\rm I} = \frac{1}{8}, ~\gamma_{\rm I} = \frac{7}{4},~\text{and}~\nu = 1.
    \label{eq50}
\end{equation}
The subscript $\rm I$ distinguishes $\beta_{\rm I}$ from the inverse strategic noise parameter $\beta$. At criticality, Eqs.~(\ref{eq49a}) and \eqref{eq49b} give $M \sim L^{-1/8}$ and $\chi_{|\mu|} \sim L^{7/4}$, respectively, while the Binder curves for a varying $L$ approach a common crossing up to finite-size corrections \cite{ref37, ref38}.

When $\kappa > \kappa_c$ and $\mathfrak{H} = 0$, the system has two symmetry related phases $\pm \mu_0$, where $\mu_0 = M_{L\rightarrow\infty}(\kappa) > 0$. For a finite volume system that samples both sectors, we have $\langle \mu \rangle \simeq 0$ and $\langle \mu^2 \rangle \rightarrow \mu_0^2$. Hence, we have,
\begin{equation}
    \chi_{\mu} (\mathfrak{H} = 0, L) = \beta L^2 (\langle \mu^2 \rangle - \langle \mu \rangle^2) \simeq \beta \mu_0^2 L^2.
    \label{eq51}
\end{equation}
In general, $L^d$ (in our case, $d=2$) volume scaling is the standard finite-size behavior of a two-phase coexistence point in $d$-dimensions \cite{ref39, ref40}. When we consider the disordered regime, i.e., $\kappa < \kappa_c$, the correlation length remains finite at $\mathfrak{H} = 0$ which leads the susceptibility to a thermodynamic value that is independent of $L$ and also $\langle |\mu |\rangle \sim N^{-1/2} = L^{-1}$. For our study, we consider these \textit{three} regimes for diagnostic purposes in Sec.~\ref{sec5}: $L^2$ scaling for ordered coexistence, $L^{7/4}$ scaling at the continuous critical point, and a finite susceptibility value for the disordered case (above the critical point). For arbitrary (\textit{real}) $\mathfrak{H} \neq 0$, we do not have any closed form analytical solution for the \textit{2D} Ising model, so we will resort to numerical simulations \cite{ref35}.

\subsection{\label{subsec3d}Channel $\rightarrow$ criticality criteria \& exact critical points}
If we consider a channel \O, then the critical point (short range) on the neutral-strategy branch must satisfy the following independent conditions,
\begin{equation}
    \mathfrak{H}_{\text{\O}} (\Gamma_c , p) = 0,~~\beta\mathfrak{J}_{\text{\O}} (\Gamma_c, p) = \kappa_c, ~~ \mathfrak{J}_{\text{\O}}> 0.
    \label{eq52}
\end{equation}

\begin{table*}[t]
\renewcommand{\arraystretch}{1.25}
\begin{ruledtabular}
\begin{tabular}{llll}
\textbf{channel} & $\mathfrak{H}=0$ branch & coupling on branch & $\beta=1$ behavior\\
\hline
phase damping & Eq.~\eqref{eq41} & $\frac{3}{4}$ & coexistence for full branch\\
depolarizing & Eq.~\eqref{eq41},~$p<1$ & $\frac{3}{4}(1-p)^2$ & critical point at $p_*$\\
AD, \textit{resource} & Eq.~\eqref{eq43} & Table~\ref{table1} & coexistence for full branch\\
AD, \textit{post-strategy} & Eq.~\eqref{eq45}, $p\le1/3$ & $\frac34(1-p)^2$ & critical point at $p_*$; branch ends at $\frac{1}{3}$\\
\end{tabular}
\end{ruledtabular}
\caption{Structure of $\mathfrak{H}=0$ branches at $\beta=1$. \textit{Coexistence} means $\beta \mathfrak{J}>\kappa_c$; \textit{crossover} means $\beta \mathfrak{J}<\kappa_c$. At $p=p_*$ the field-driven coexistence line terminates at the square-lattice Ising critical point (see, Fig.~\ref{fig2}).}
\label{table2}

\end{table*}
\begin{figure*}[t]
\centering
\begin{subfigure}[t]{0.32\textwidth}
    \centering
    \includegraphics[width=\linewidth]{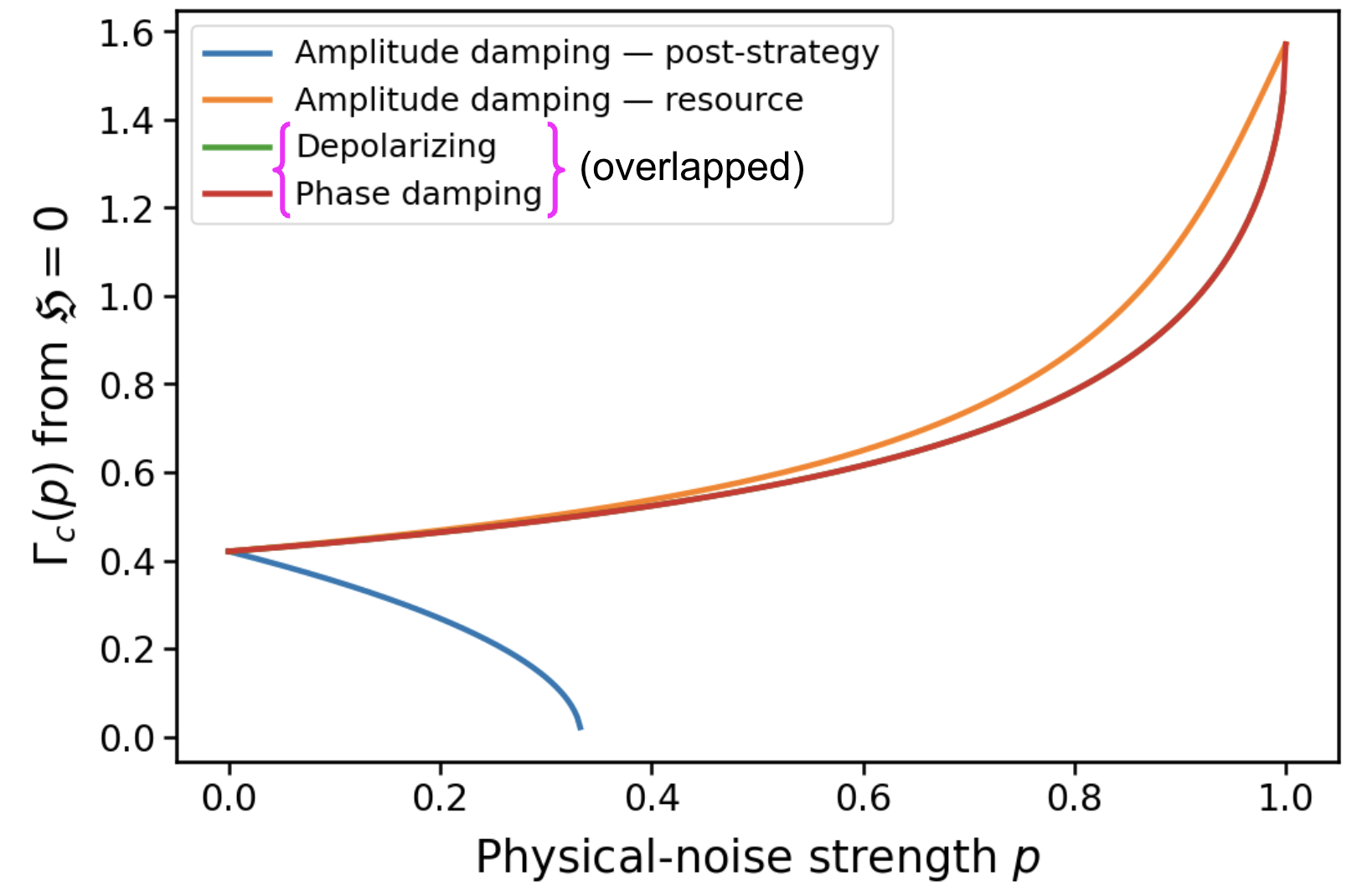}
    \caption{}
    \label{fig2a}
\end{subfigure}
\hfill
\begin{subfigure}[t]{0.32\textwidth}
    \centering
    \includegraphics[width=\linewidth]{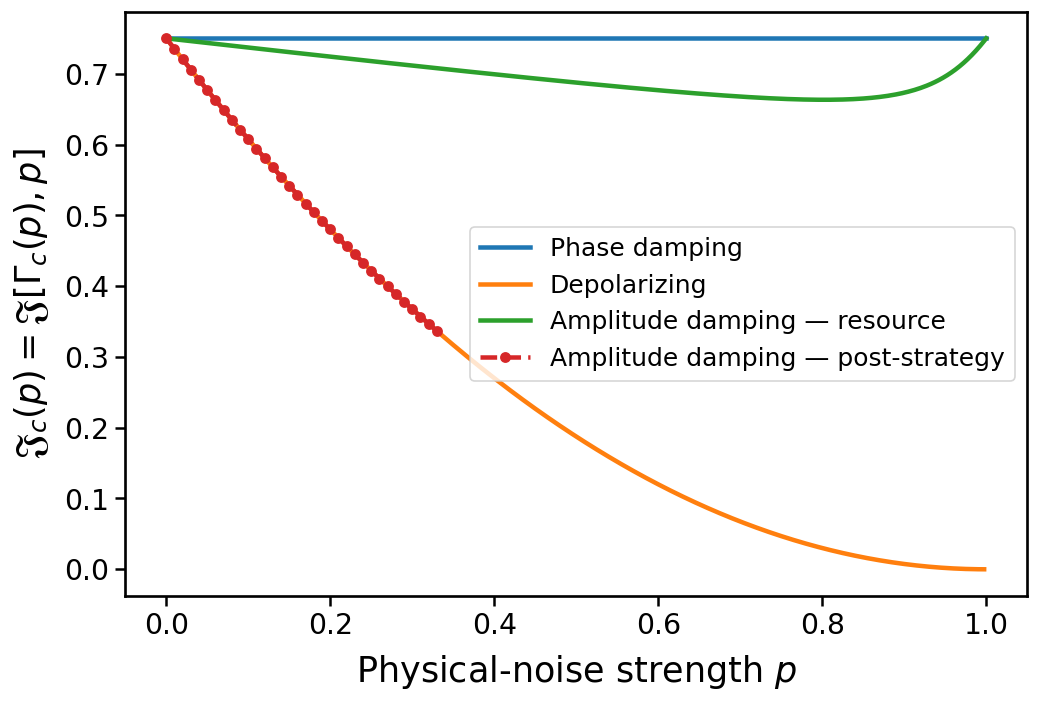}
    \caption{}
    \label{fig2b}
\end{subfigure}
\hfill
\begin{subfigure}[t]{0.32\textwidth}
    \centering
    \includegraphics[width=\linewidth]{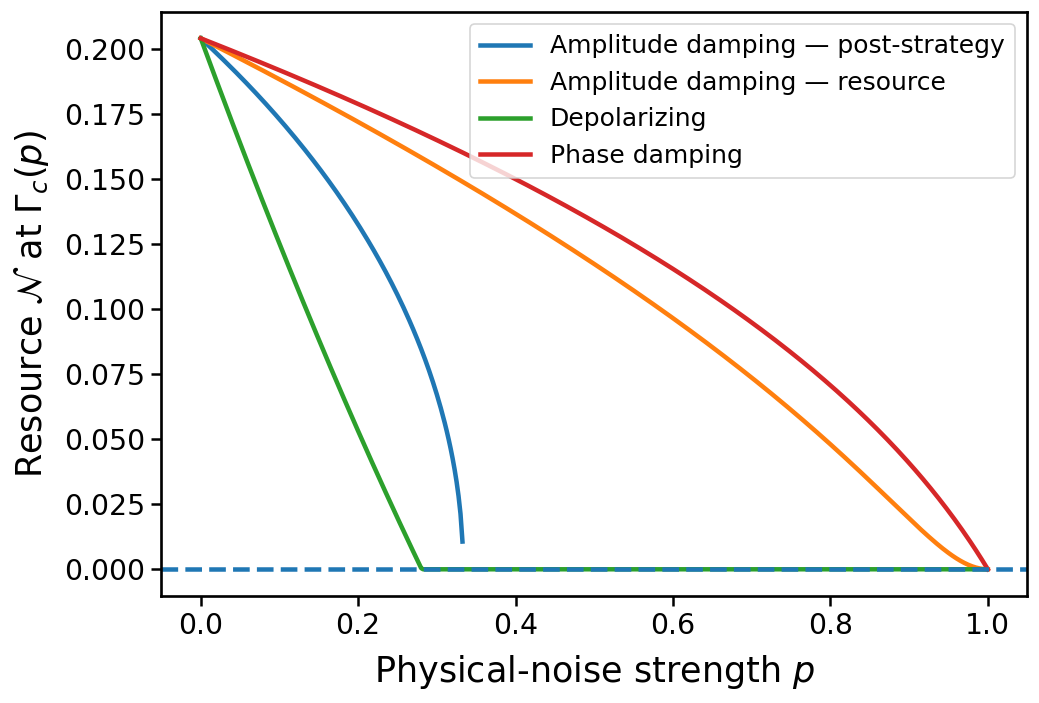}
    \caption{}
    \label{fig2c}
\end{subfigure}
\caption{Exact channel deformation along the microscopic neutrality line. \textbf{(a)} Zero-field entanglement $\Gamma_c(p)$. Phase damping and depolarization share the same $\mathfrak{H}=0$ curve and are therefore shown as an overlapped branch. The post-strategy amplitude damping branch terminates at $p=1/3$. \textbf{(b)} Effective interaction $\mathfrak{J}_c(p)=\mathfrak{J}[\Gamma_c(p),p]$. Depolarization and post-strategy amplitude damping obey the same coupling law, $\mathfrak{J}_c=\frac34(1-p)^2$, wherever the post strategy $\mathfrak{H}=0$ branch exists. \textbf{(c)} Negativity of the \textit{EWL} resource immediately before the strategic layer. For resource-noise channels, this is the negativity of the state \textit{after} the channel has acted. For post-strategy amplitude damping, the channel has \textit{not} yet acted at this stage, so the plotted quantity is the \textit{noiseless} resource negativity, $\frac{1}{2}\sin\Gamma_c^{(\mathcal{S})}(p)$, evaluated along the corresponding placement-dependent neutrality branch.}
\label{fig2}
\end{figure*}

The first expression in Eq.~(\ref{eq52}) makes the two strategic orientations within the potential \textit{unbiased}, while the second expression identifies if the neutral point is either ordered, disordered, or critical. This forms the main theoretical basis of the paper.

The exact critical inverse strategic noise along a zero-field branch is given as,
\begin{equation}
  \beta_c(p)=\frac{\kappa_c}{\mathfrak{J}[\Gamma_c(p),p]}.
  \label{eq53}
\end{equation}
For the channels with simple expression for $\mathfrak{J}$,
\begin{gather}
  \beta_{c,~ph} =\frac{4\kappa_c}{3},~~
  \beta_{c,~dep}(p) =\frac{4\kappa_c}{3(1-p)^2},\\
  \beta_{c,~ad}^{(\mathcal{S})}(p) =\frac{4\kappa_c}{3(1-p)^2},~~ p\in \bigg[0, \frac{1}{3}\bigg],
  \label{eq54}
\end{gather}
while $\beta_{c,~ad}^{(\mathcal{R})}(p)$ can be determined by substituting Eq.~\eqref{eq43} into the resource-damping $\mathfrak{J}$ of Table~\ref{table1}.

For depolarizing noise at fixed $\beta$, Eq.~\eqref{eq52} gives the exact critical point as,
\begin{equation}
  p_*(\beta)=1-\sqrt{\frac{4\kappa_c}{3\beta}},
  \label{eq55}
\end{equation}
given that a physical solution exists. Since $0\le p_*<1$, the required strategic-noise regime is,
\begin{equation}
  \beta\ge \beta_{\min}\equiv\frac{4\kappa_c}{3}
  \approx 0.587582\ldots.
  \label{eq56}
\end{equation}
At $\beta=\beta_{\min}$, the critical point begins at $p_*=0$. For $\beta<\beta_{\min}$ the population is already on the disordered side of the square-lattice transition at $p=0$, and depolarization can only reduce $\mathfrak{J}$ further, so no decoherence-driven crossing occurs. For $\beta>\beta_{\min}$, $0<p_*<1$ and $p_*\to 1$ as $\beta\to\infty$. We use $\beta=1$ in the simulations since it acts as a particular representation across a broader $(\beta,p)$ structure. At $\beta=1$,
\begin{equation}
  p_*\approx 0.233460\ldots,
  ~~
  \Gamma_*\approx 0.472247\ldots.
  \label{eq57}
\end{equation}
Since \textit{post-strategy} amplitude damping has the same $\mathfrak{J}=\frac{3}{4}(1-p)^2$, its $\mathfrak{H}=0$ branch reaches the same square-lattice critical condition at $p_*$ when $\beta=1$, but at a different value of $\Gamma_c$. The branch then remains \textit{neutral} but thermodynamically disordered for $p_*<p\le \frac{1}{3}$, after which its nontrivial $\mathfrak{H}=0$ part disappears. Meanwhile, phase damping has $\beta \mathfrak{J}=0.75>\kappa_c~ \forall~ p$ at $\beta=1$, and the exact resource-amplitude branch likewise remains on the ordered side of $\kappa_c$ across its entire interval.

These results are summarized in Table~\ref{table2}. In particular, phase damping and depolarization satisfy the {same} Eq.~\eqref{eq41} for microscopic neutrality, yet their collective behavior differs because $\mathfrak{J}$ is different.

For $\beta \mathfrak{J}>\kappa_c$, $\mathfrak{H}=0$ is a field-driven \textit{first}-order coexistence line between the two magnetized strategic orientations. At $\beta \mathfrak{J} = \kappa_c$ it terminates at the \textit{2D} Ising critical point. For $\beta \mathfrak{J}< \kappa_c$, changing the sign of $\mathfrak{H}$ results in a smooth crossover. From this we can realize that the microscopic quantum channel controls the system's location within the standard Ising phase structure.

\subsection{\label{subsec3e}Microscopic resource entanglement}
We should realize that the \textit{EWL} resource is generally a two-qubit mixed state after the action of physical noise, while the lattice model is classical in nature. For microscopic diagnostic purposes to keep these two things separate, we use entanglement \textit{negativity}. For a bipartite state $\varrho$, the \textit{negativity} is defined as \cite{ref41},
\begin{equation}
  \mathcal{N}(\varrho)=\frac{\|\varrho^{T_B}\|_1 - 1}{2},
  \label{eq58}
\end{equation}
where $T_B$ indicates partial transpose and $\|\cdot\|_1$ is the trace norm. For the state immediately after \emph{resource} noise according to the current channel conventions,
\begin{gather}
  \mathcal{N}_{ph}(\Gamma,p) =\frac{1-p}{2}\sin\Gamma, \label{eq59a}\\
  \mathcal{N}_{dep}(\Gamma,p) =\max\left[0,\frac{(1-p)^2}{2}\sin\Gamma-\frac{p(2-p)}{4}\right] \label{eq59b},\\
  \mathcal{N}_{ ad}(\Gamma,p) =\frac{1-p}{2}\left[\sin\Gamma-p(1-\cos\Gamma)\right].
  \label{eq59c}
\end{gather}
The resource density matrices and partial transpose eigenvalues leading to Eqs.~\eqref{eq59a}-\eqref{eq59c} are given in Appendix~\ref{appendixC}.

In the depolarizing zero-field branch, the \textit{resource} becomes separable at $p\approx 0.280719$, after the $\beta=1$ collective critical point $p_*$ has already been crossed. This ordering of thresholds should \textit{not} be interpreted as a quantum many-body phase persisting after the entanglement vanishes. Eq.~\eqref{eq34} is an ordinary classical Gibbs state; $\mathcal{N}$ diagnoses only the microscopic two-qubit resource from which the effective $(\mathfrak{J}, \mathfrak{H})$ were constructed. The distinction between microscopic quantum correlations and collective classical correlations is important in thermodynamic limit quantum-game setups~\cite{ref13, ref14}.

\section{\label{sec4}Monte Carlo Technique}
We simulate Eq.~\eqref{eq33} on periodic square lattices of size $N=L^2$ with,
\begin{equation}
  L\in\{16,24,32,48,64\}.
  \label{eq60}
\end{equation}
The implementation follows the standard equilibrium importance-sampling practice for Ising systems~\cite{ref39, ref42, ref43}. Since every lattice size used here is \textit{even}, the periodic square lattice is \textit{bipartite} and can be updated in two sub-lattices without simultaneous N.N. conflicts. For a proposed local flip $\mathfrak{s}_i\to-\mathfrak{s}_i$, Eq.~\eqref{eq33} gives us,
\begin{equation}
   \partial E_i=2\mathfrak{s}_i\left[\mathfrak{J}\sum_{j\in\mathcal N(i)}\mathfrak{s}_j+4\mathfrak{H}\right],
  \label{eq61}
\end{equation}
which is \textit{accepted} with the Metropolis probability,
\begin{equation}
  \text{Prob}_{\rm acc}=\min\left[1,e^{-\beta\partial E_i}\right].
  \label{eq62}
\end{equation}
The total energy is evaluated by counting only one \textit{horizontal} and one \textit{vertical} directed bond per site, so each undirected N.N. edge is included exactly once.

Near $\mathfrak{H}=0$ and for $\kappa > \kappa_c$, purely local dynamics can remain ``\textit{trapped}" for long periods in one of the two magnetized sectors. Hence, we supplement the local sweeps by proposing a \textit{global} inversion $\mathfrak{s}_i\to-\mathfrak{s}_i$ with probability $0.25$ per sweep. The bond term is invariant under this, while the field contribution changes by,
\begin{equation}
  \partial E_{\rm global}=8\mathfrak{H}\sum_i \mathfrak{s}_i.
  \label{eq63}
\end{equation}

This is accepted via the same Metropolis rule, so the balance with respect to Eq.~\eqref{eq34} is preserved. At $\mathfrak{H}=0$, $\partial  E_{\rm global}=0$ and this accelerates sampling of the two symmetry-related sectors; it does not change the target distribution. When the system is not near the neutral point, an ordered configuration aligned with the field has $\partial E_{\rm global}\simeq8|\mathfrak{H}|\,|\mu|L^2$, so the acceptance of the global move becomes exponentially small with system area. Hence, it contributes appreciably only near $\mathfrak{H}=0$ in the $\Gamma$ scans; ordinary local Metropolis updates provide the sampling away from that region.

The measured magnetization per site and its absolute value are given as,
\begin{equation}
  \mu = \frac{1}{N}\sum_i \mathfrak{s}_i,~~
  M=\langle{|\mu|}\rangle.
  \label{eq64}
\end{equation} 
We consider both the signed and absolute-magnetization susceptibilities,
\begin{gather}
  \chi_\mu =\beta N(\langle{\mu^2}\rangle-\langle{\mu}\rangle^2),\\
  \chi_{|\mu|} =\beta N(\langle{\mu^2}\rangle-\langle{|\mu|\rangle}^2),
  \label{eq65}
\end{gather}
and the \textit{fourth}-order Binder cumulant is given as~\cite{ref37},
\begin{equation}
  U_4=1-\frac{\langle{\mu^4}\rangle}{3\langle{\mu^2}\rangle^2}.
  \label{eq66}
\end{equation}
\begin{table*}[t]
\begin{ruledtabular}
\begin{tabular}{lccccc}
\textit{Dataset} & \textit{Seeds} & \textit{Thermalization sweeps} & \textit{Measure} & \textit{Sample every} & \textit{Jobs}\\
\hline
zero-field $p$ scan & 4 & 12000 & 40000 & 5 & 460\\
exact-$p_*$ validation & 6 & 25000 & 100000 & 5 & 30\\
depolarizing $\Gamma$ scan, each $p$ & 6 & 20000 & 80000 & 5 & 1470\\
amplitude $\Gamma$ scan, each placement & 6 & 20000 & 80000 & 5 & 1470\\
$P(\mu)$, each $p$ at $L=64$ & 6 & 25000 & 120000 & 5 & 6\\
\end{tabular}
\end{ruledtabular}
\caption{Monte Carlo datasets used in the figures. The $\Gamma$ scans contain 49 points over $\Gamma_c\pm0.08$.  ``\textit{Measure}'' denotes the number of measurement sweeps before sub-sampling. ``\textit{Sample every}" denotes one sample being recorded every $n~(=5)$ sweeps. ``\textit{Jobs}" denotes the total number of parameter-point simulations.}
\label{table3}
\end{table*}
We intentionally consider the two different kinds of susceptibilities. $M$ and $\chi_{|\mu|}$ avoid the cancellation of the order parameter, and we use them for continuous critical point finite-size scaling. The signed $\chi_\mu$ is instead the natural diagnostic of the field-driven coexistence since it retains fluctuations between the $+\mu_0$ and $-\mu_0$ sectors, leading to the $L^2$ behavior in Eq.~\eqref{eq51}.

\begin{figure*}[t]
\centering
\begin{subfigure}[t]{0.32\textwidth}
    \centering
    \includegraphics[width=\linewidth]{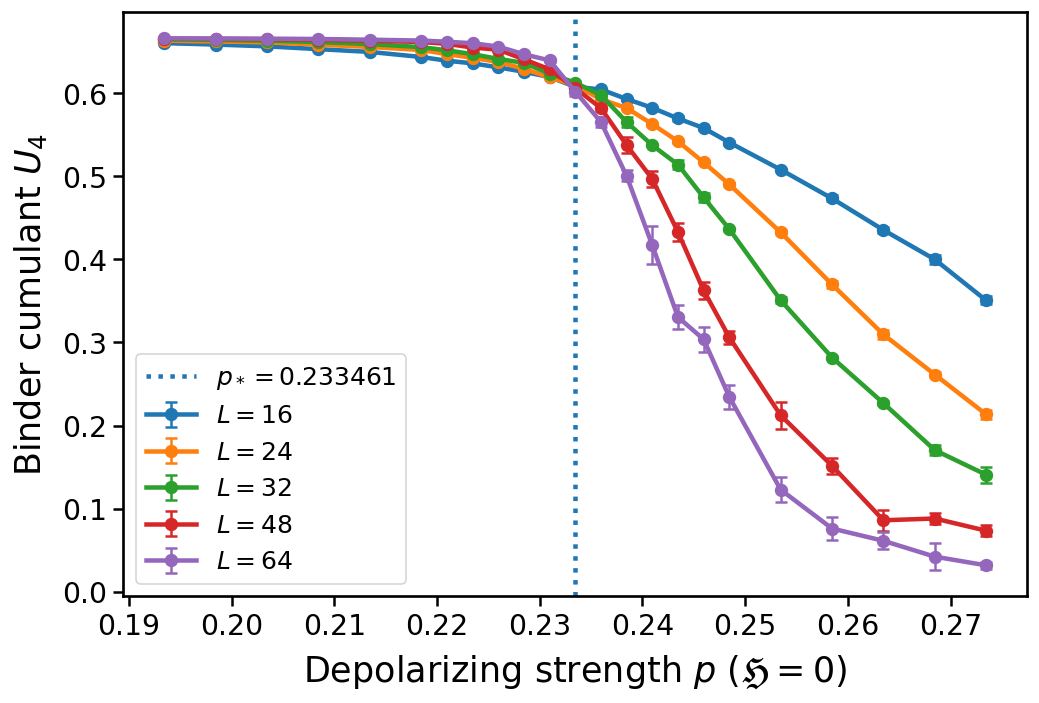}
    \caption{The dotted line marks the analytical $p_{*} \approx 0.233460\ldots$.}
    \label{fig3a}
\end{subfigure}
\hfill
\begin{subfigure}[t]{0.32\textwidth}
    \centering
    \includegraphics[width=\linewidth]{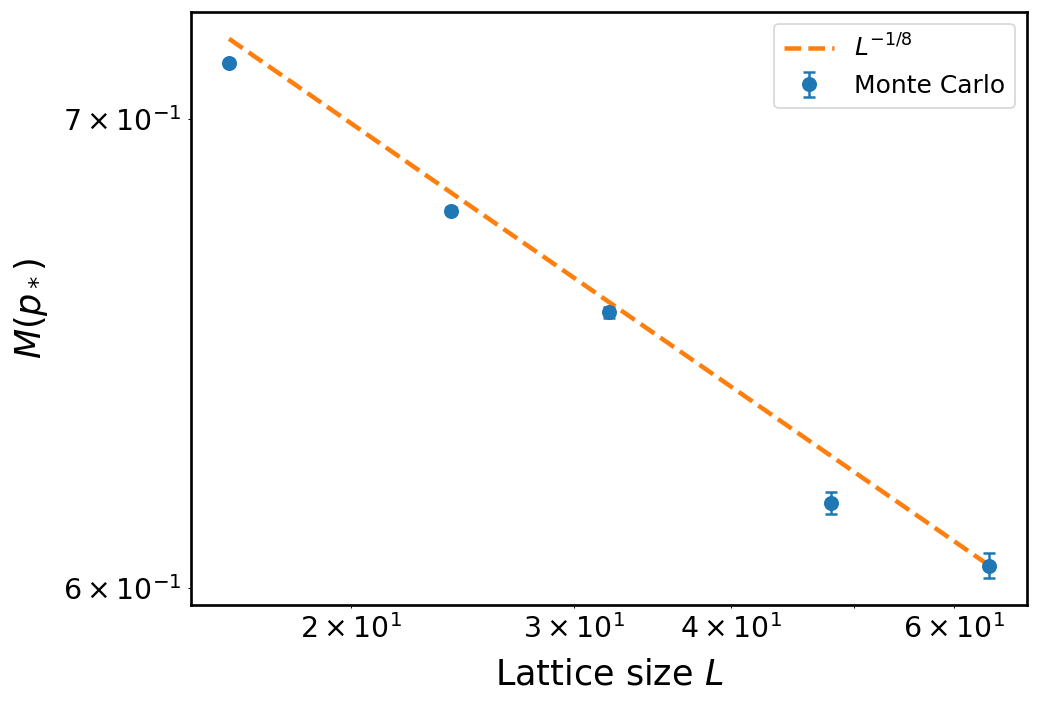}
    \caption{Finite size $M$ scaling at $p_{*}$}
    \label{fig3b}
\end{subfigure}
\hfill
\begin{subfigure}[t]{0.32\textwidth}
    \centering
    \includegraphics[width=\linewidth]{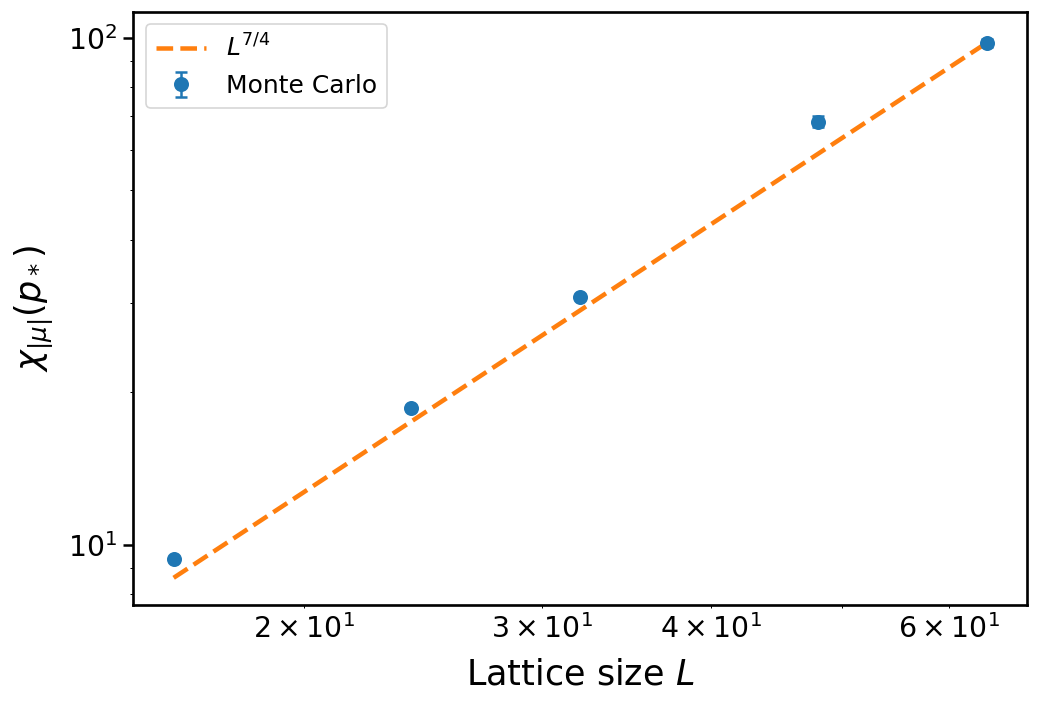}
    \caption{Finite size $\chi_{|\mu|}$ scaling at $p_{*}$}
    \label{fig3c}
\end{subfigure}
\caption{Monte Carlo validation of the depolarizing critical point at $\beta=1$. \textbf{(a)} Binder cumulant along the exact microscopic $\mathfrak{H} = 0$ line. \textbf{(b,c)} High-statistics finite-size scaling performed directly at $p_{*}$. The dashed references are the exact \textit{2D} Ising powers $L^{-1/8}$ and $L^{7/4}$.}
\label{fig3}
\end{figure*}

Table~\ref{table3} lists the production datasets used in the simulation figures. Here, each parameter point is simulated with independent random number seeds (we call this a \textit{replica}). The moments $\langle\mu\rangle$, $\langle |\mu |\rangle$, $\langle \mu^2 \rangle$, and $\langle\mu^4 \rangle$ are averaged over the independent \textit{replicas} before the nonlinear observables are constructed. Their uncertainties are estimated using a \textit{delete-one-replica jackknife}. Binder crossing errors are obtained separately by \textit{bootstrap resampling} of the replicas. These are standard error-estimation procedures for Monte Carlo data \cite{ref39, ref43}.

Within each replica, we also estimate the integrated autocorrelation time of $|\mu|,~\mu^2$ and the energy from the normalized autocorrelation function \cite{ref39, ref43}, 
\begin{equation}
    \tau_{\rm int}=\frac{1}{2}+\sum_{t=1}^{W}\varrho(t),
\label{eq67}
\end{equation}
where the sum is truncated using a positive, self-consistent window $W$. The corresponding effective sample size for $|\mu|$ is estimated as,
\begin{equation}
    n_{\rm eff}\simeq\frac{n_{\rm samp}}{2\tau_{\rm int}}.
\label{eq68}
\end{equation}

We also calculate non-overlapping block averages of $|\mu|$ using a block length of approximately $4\tau_{\rm int}$ as an additional convergence check within each replica. Since local Metropolis updates exhibit critical slowing down near a continuous phase transition, these diagnostics are used mainly to check that the simulations are sufficiently long, rather than to replace the error bars obtained from the independent replicas~\cite{ref39, ref43}. Detailed results for the zero-field runs are given in Appendix~\ref{appendixE}.

Finally, we independently verify the analytical \textit{EWL}-to-Ising mapping using a direct $4\times4$ density matrix calculation. The \textit{EWL} circuit and Kraus maps are implemented numerically, the four $\mathsf{Q/D}$ payoffs are calculated, and the resulting $\mathfrak{J}$ and $\mathfrak{H}$ are obtained from Eq.~\eqref{eq27}. Tests at randomly chosen $(\Gamma,p)$ points agree with the analytical expressions and the independent negativity formulas to numerical precision. Further details are given in Appendix~\ref{appendixD}. 

OpenAI GPT-5.6 Sol was used to assist with code drafting and debugging. The physical model, simulation setup, and analysis were determined by the author. All final code and numerical results were run, checked, and verified by the author.

\section{\label{sec5}Results \& Discussion}

\subsection{\label{subsec5a}Zero-field results for depolarization-driven critical point}
\begin{figure*}[t]
\centering
\begin{subfigure}[t]{0.32\textwidth}
    \centering
    \includegraphics[width=\linewidth]{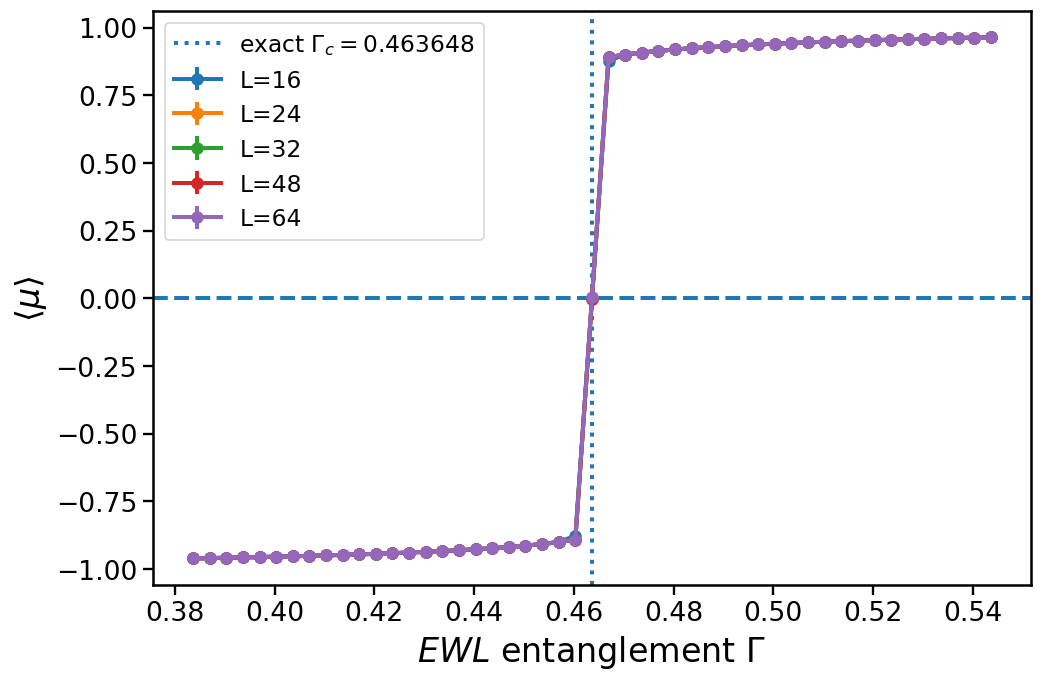}
    \caption{}
    \label{fig4a}
\end{subfigure}
\hfill
\begin{subfigure}[t]{0.32\textwidth}
    \centering
    \includegraphics[width=\linewidth]{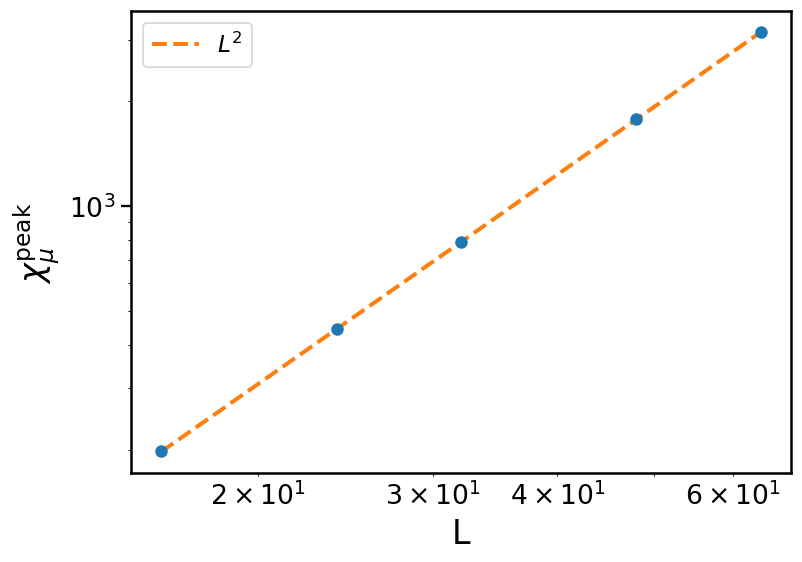}
    \caption{}
    \label{fig4b}
\end{subfigure}
\hfill
\begin{subfigure}[t]{0.32\textwidth}
    \centering
    \includegraphics[width=\linewidth]{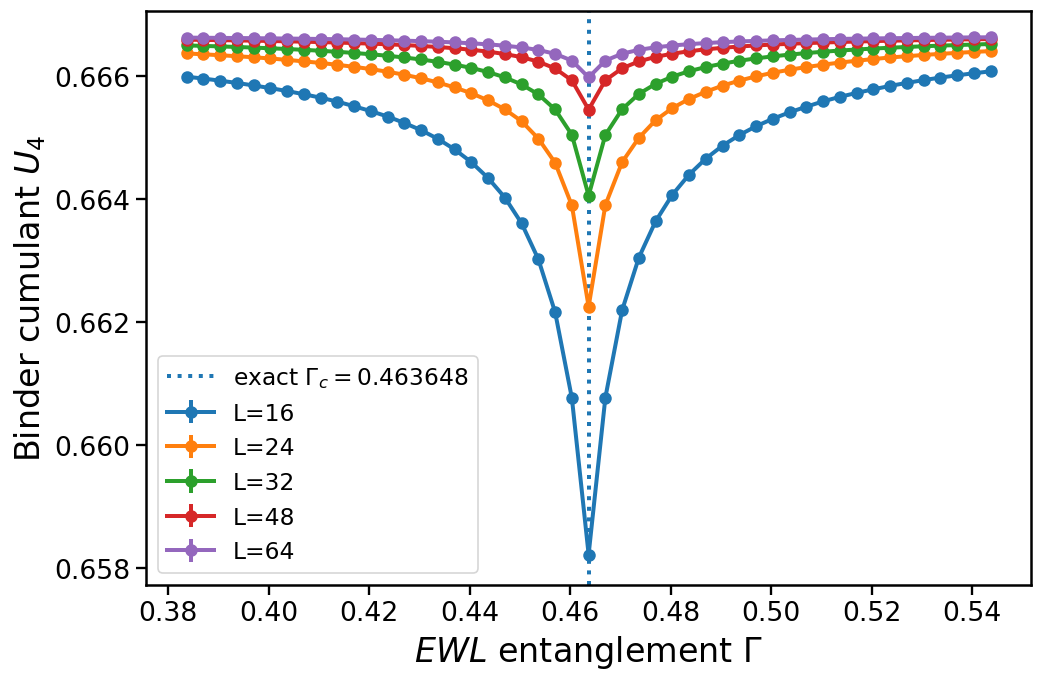}
    \caption{}
    \label{fig4e}
\end{subfigure}
\hfill
\begin{subfigure}[t]{0.32\textwidth}
    \centering
    \includegraphics[width=\linewidth]{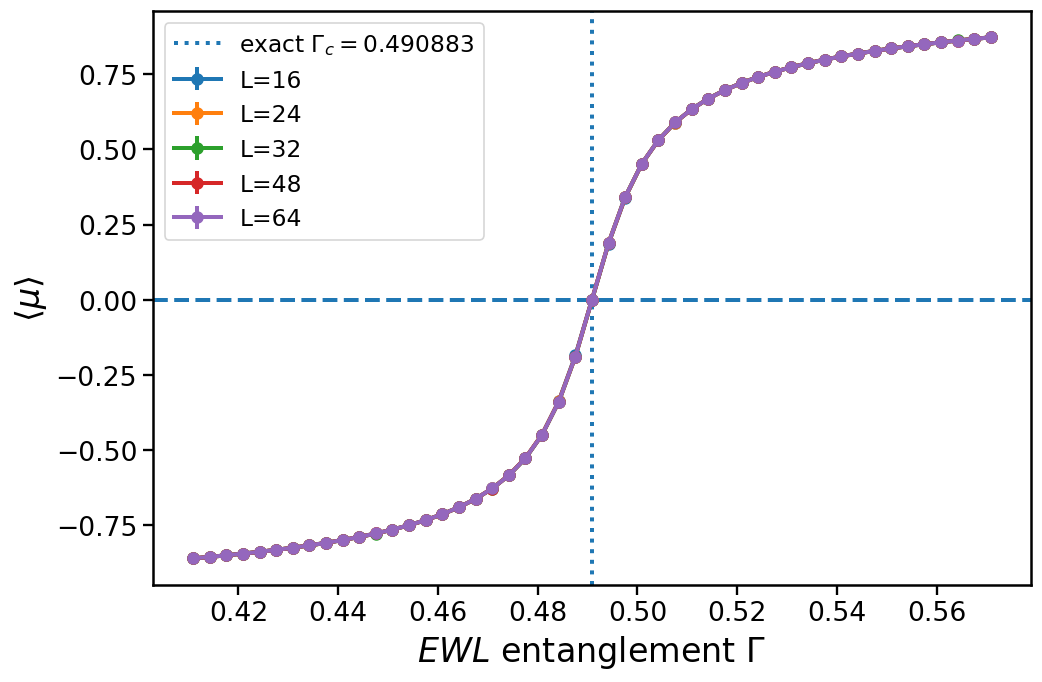}
    \caption{}
    \label{fig4c}
\end{subfigure}
\hfill
\begin{subfigure}[t]{0.32\textwidth}
    \centering
    \includegraphics[width=\linewidth]{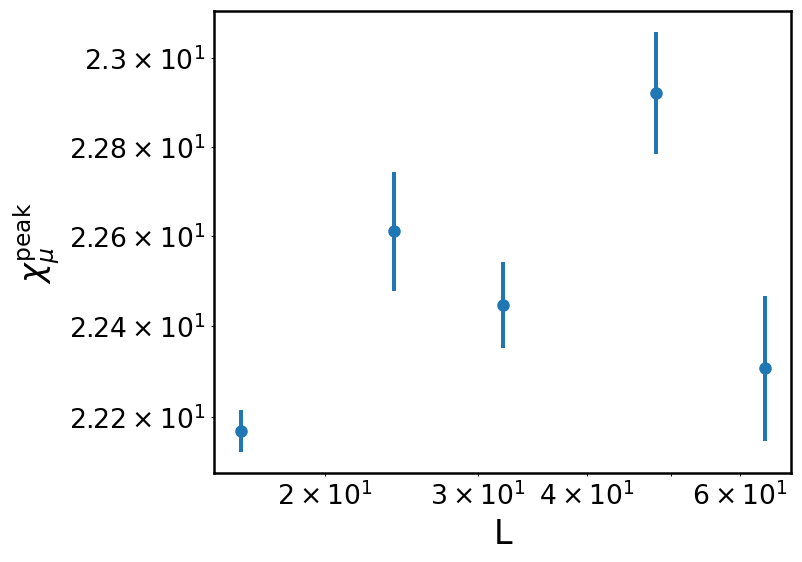}
    \caption{}
    \label{fig4d}
\end{subfigure}
\hfill
\begin{subfigure}[t]{0.32\textwidth}
    \centering
    \includegraphics[width=\linewidth]{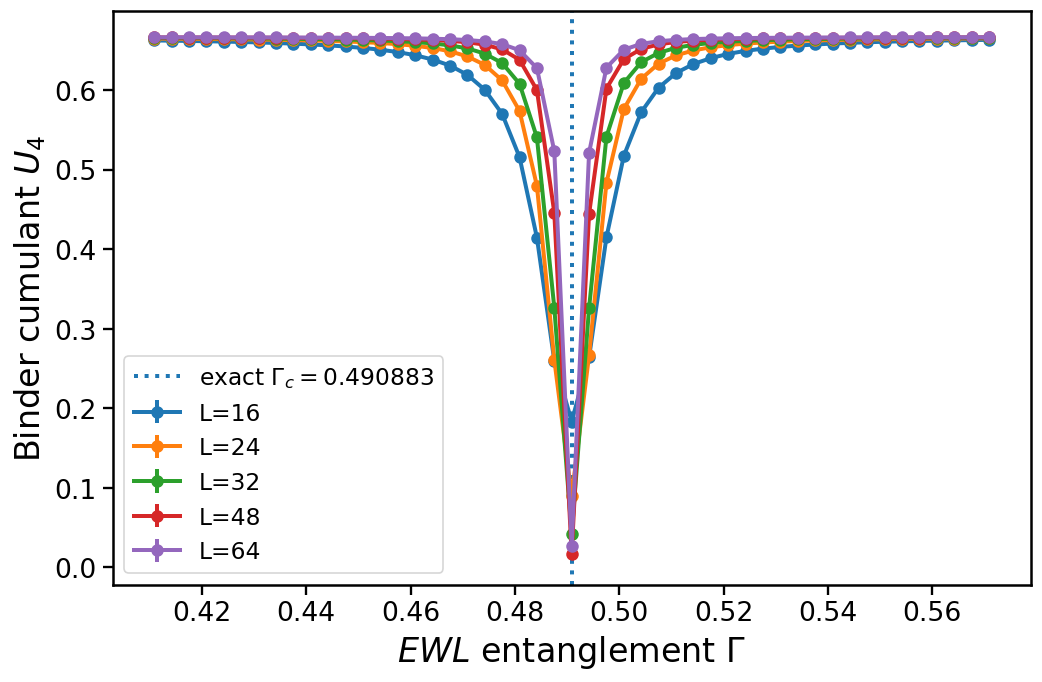}
    \caption{}
    \label{fig4f}
\end{subfigure}
\caption{Fixed-$\beta$ depolarizing scans through the exact $\mathfrak{H}=0$ point, including $\mathfrak{H}\neq 0$ data on both sides. In panels \textbf{(a,d)}, points to the left of the vertical dotted line have $\mathfrak{H}<0$ and favor $\mathsf{D}$, the dotted line itself is $\mathfrak{H}=0$, and points to the right have $\mathfrak{H}>0$ and favor $\mathsf{Q}$. \textbf{(a--c)} For $p=0.20<p_{*}$, $\langle\mu\rangle$ switches sharply from a
$\mathsf{D}$-dominated state ($\langle \mu\rangle<0$) to a $\mathsf{Q}$-dominated state ($\langle \mu \rangle>0$), and $\chi_\mu^{\rm peak}$ follows $L^2$. \textbf{(d--f)} For $p=0.30>p_{*}$, the same $\mathsf{D}\to\mathsf{Q}$ change is a smooth crossover and the susceptibility maximum shows no systematic growth with $L$. The Binder cumulant gives us a complementary distinction: for $p=0.20$, $U_4$ near $\mathfrak{H}=0$ remains close to the ordered value $2/3$ as $L$ increases, whereas for $p=0.30$ it develops a minimum approaching $U_4=0$, indicating a disordered state at $\mathfrak{H}=0$.}
\label{fig4}
\end{figure*}
We first use Monte Carlo on the exactly solvable $\mathfrak{H}=0$ line as a validation of the noise channel-to-Ising mapping. Eq.~\eqref{eq57} predicts $p_*$ analytically, without taking into consideration the simulation data. The finite-size Binder curves cross in its vicinity, as seen in Fig.~\ref{fig3a}. Pairwise \textit{linear} interpolation estimates are as follows,
\begin{center}
\begin{tabular}{ccc}
\toprule
$(L_1,L_2)$ & $p_{\times}$ & bootstrap $16$--$84\%$ interval\\
\midrule
$(16,24)$ & 0.233800 & [0.232120, 0.234109]\\
$(24,32)$ & 0.236541 & [0.234719, 0.236931]\\
$(32,48)$ & 0.232071 & [0.231390, 0.232609]\\
$(48,64)$ & 0.232761 & [0.232215, 0.233649]\\
\bottomrule
\end{tabular}
\label{tab1}
\end{center}
The smaller-size Binder crossings show a non-monotonic drift, which we report directly rather than using an extrapolation. The crossing for the largest pair of system sizes is statistically consistent with the exact value $p_{*}$. The bootstrap intervals listed above reflect the statistical uncertainty of each finite-size crossing $p_{\times}(L_1,L_2)$. Due to finite-size corrections, these pseudo-critical crossings need not coincide exactly with the thermodynamic critical point $p_{*}$.

A cleaner validation is obtained by simulating directly at the analytically predicted critical point. The Binder crossing strategy and power-law tests are standard finite-size diagnostics of Ising criticality~\cite{ref37, ref38, ref39}. For the \textit{2D} Ising universality class,
\begin{equation}
  M(p_*,L)\sim L^{-\frac{1}{8}},
  ~~
  \chi_{|\mu|}(p_*,L)\sim L^{\frac{7}{4}}.
  \label{eq69}
\end{equation}
For weighted log--log fits to the high-statistics runs, we get,
\begin{gather}
  M(p_*,L) \sim L^{-0.1223\pm0.0040},~
  \chi_{|\mu|}(p_*,L) \sim L^{1.717\pm0.027}.
  \label{eq70}
\end{gather}

The corresponding bootstrap means are $-0.1235$ and $1.729$. Within finite-size corrections, the results are consistent with $-\frac{1}{8}$ and $\frac{7}{4}$ (see, Figs.~\ref{fig3b}, \ref{fig3c}). Since the zero-field square-lattice critical point is known exactly, these simulations are used as an implementation and universality check rather than as the primary determination of the value of $p_{*}$.

%_____________________________________

%_______________________________________

\subsection{\label{subsec5b}Critical point: coexistence \& crossover}
\begin{figure*}[t]
\centering
\includegraphics[width=0.47\linewidth]{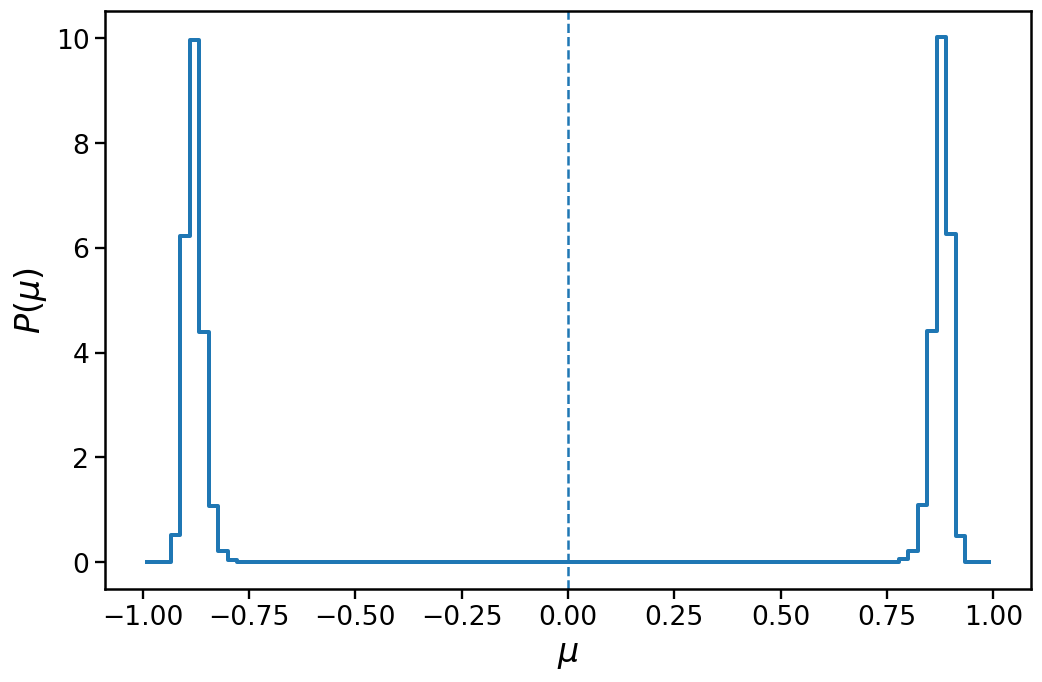}
\includegraphics[width=0.47\linewidth]{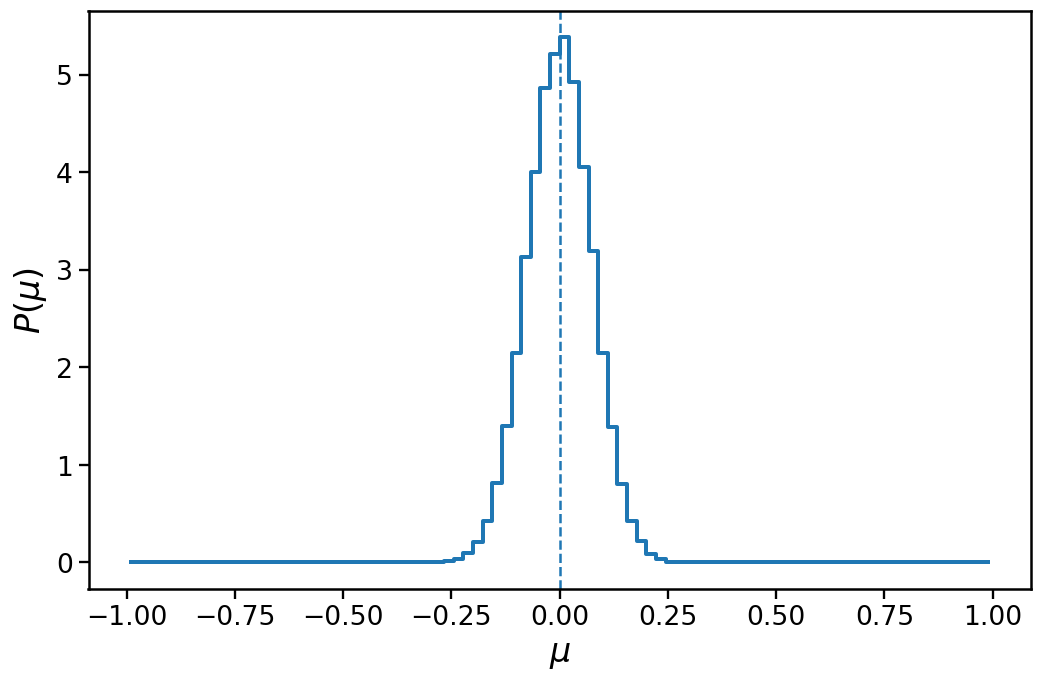}
\caption{Magnetization distributions at $L=64$, $\beta=1$, and $\mathfrak{H}=0$ point. \textbf{(a)} $p=0.20$: bimodal coexistence distribution. \textbf{(b)} $p=0.30$: single disordered distribution around \textit{zero} magnetization.}
\label{fig5}
\end{figure*}
The physical meaning of the critical point becomes more evident in the nonzero-field case obtained by scanning the \textit{EWL} entanglement $\Gamma$ through $\mathfrak{H}=0$ at two fixed depolarizing strengths. For the depolarizing channel, $\mathfrak{H}(\Gamma,p)$ increases through zero on the physical branch, so $\Gamma<\Gamma_c$ corresponds to $\mathfrak{H}<0$ and a $\mathsf{D}$-favoring field, whereas $\Gamma>\Gamma_c$ corresponds to $\mathfrak{H}>0$ and a $\mathsf{Q}$-favoring field. Only the single point $\Gamma=\Gamma_c$ has $\mathfrak{H}=0$; the surrounding area is genuinely nonzero field. At $p=0.20$,
\begin{equation}
  \Gamma_c \approx 0.463648,~~\mathfrak{J}_c=0.48>\kappa_c,
  \label{eq71}
\end{equation}

so the zero-field point lies in the \textit{ordered} regime at $\beta=1$. The signed magnetization shows a sharp turn between the two strategic phases as $L$ grows (zoom into Fig.~\ref{fig4a}). The signed susceptibility peak (see, Fig.~\ref{fig4b}) obeys,
\begin{equation}
  \chi_\mu^{\rm peak}\sim L^{1.9960\pm0.0011},
  \label{eq72}
\end{equation}
consistent with the $L^d=L^2$ volume scaling expected for two-phase coexistence in $d=2$~\cite{ref40}. At the peak, $M\simeq0.8774$ for all values of $L$ (see, Fig.~\ref{figapp4} in Appendix~\ref{appendixF}), and $\chi_\mu^{\rm peak}/L^2\to 0.770$, consistent with $\mu_0^2$ for a nearly symmetric bimodal distribution.

At $p=0.30$,
\begin{equation}
  \Gamma_c \approx 0.490883,~~
  \mathfrak{J}_c=0.3675<\kappa_c,
  \label{eq73}
\end{equation}
so $\mathfrak{H}=0$ lies in the disordered regime. The sign change in $\langle \mu \rangle$ is a smooth crossover (see, Fig.~\ref{fig4c}), and the susceptibility peak remains approximately $22$ over the full size range (see, Fig.~\ref{fig4d}). A log--log fit gives us,
\begin{equation}
  \chi_\mu^{\rm peak}\sim L^{0.017\pm0.009},
  \label{eq74}
\end{equation}
consistent with finite thermodynamic susceptibility. The value of $M$ at $\mathfrak{H}=0$ falls approximately as $L^{-1}$, as expected for a finite-correlation length \textit{2D} paramagnet (see, Fig.~\ref{figapp4} in Appendix~\ref{appendixF}).

The order-parameter distributions are shown in Fig.~\ref{fig5}. At $p=0.20$, for $L=64$ and at $\mathfrak{H}=0$, the distribution is clearly bimodal, with two narrow peaks near $\mu \approx\pm0.88$. These correspond to the coexisting $\mathsf{Q}$- and $\mathsf{D}$-dominated phases. In contrast, at $p=0.30$ the distribution has a single peak centered around $\mu = 0$, consistent with the disordered regime. Each histogram contains $144000$ recorded magnetization samples collected from \textit{six} independent runs.

\begin{figure*}[t]
\centering
\begin{subfigure}[t]{0.45\textwidth}
    \centering
    \includegraphics[width=\linewidth]{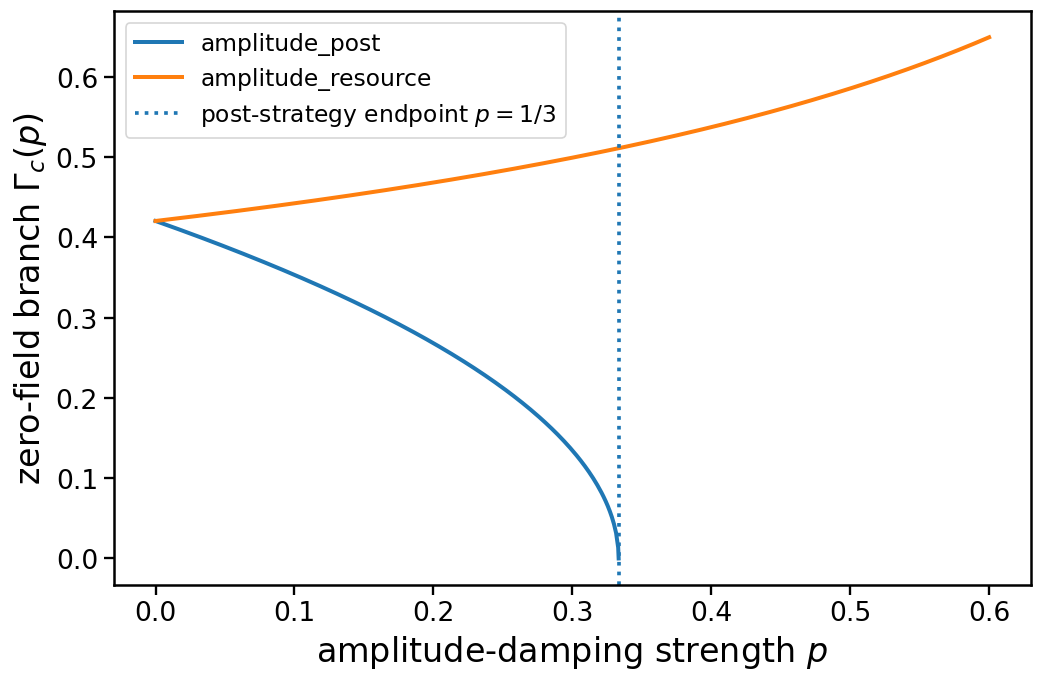}
    \caption{}
    \label{fig6a}
\end{subfigure}
\hfill
\begin{subfigure}[t]{0.45\textwidth}
    \centering
    \includegraphics[width=\linewidth]{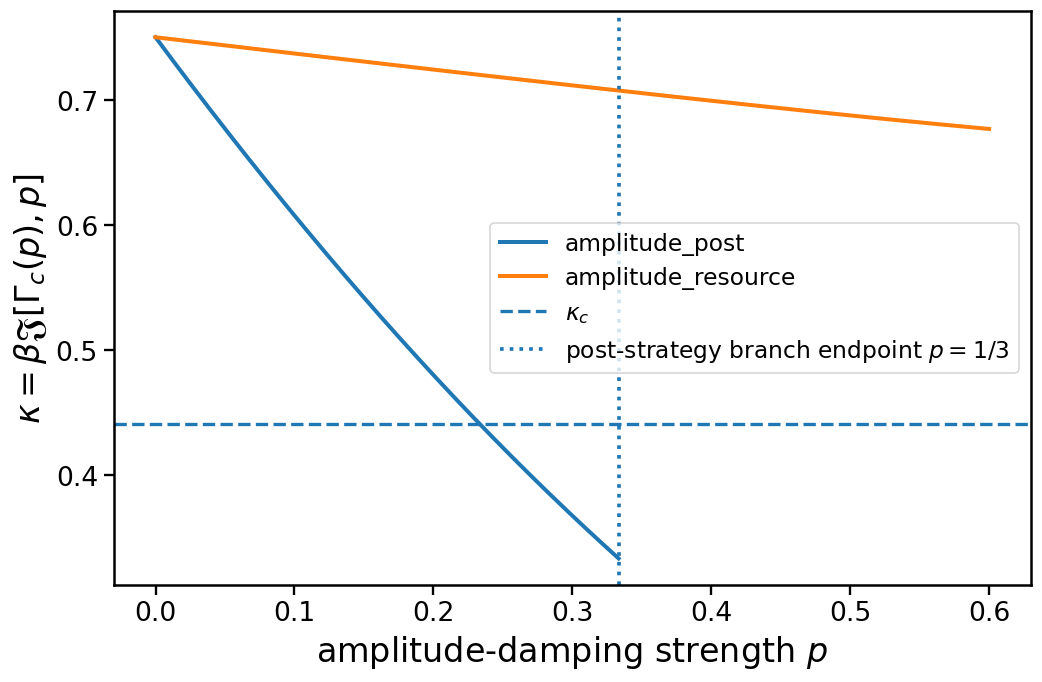}
    \caption{}
    \label{fig6b}
\end{subfigure}
\hfill
\begin{subfigure}[t]{0.45\textwidth}
    \centering
    \includegraphics[width=\linewidth]{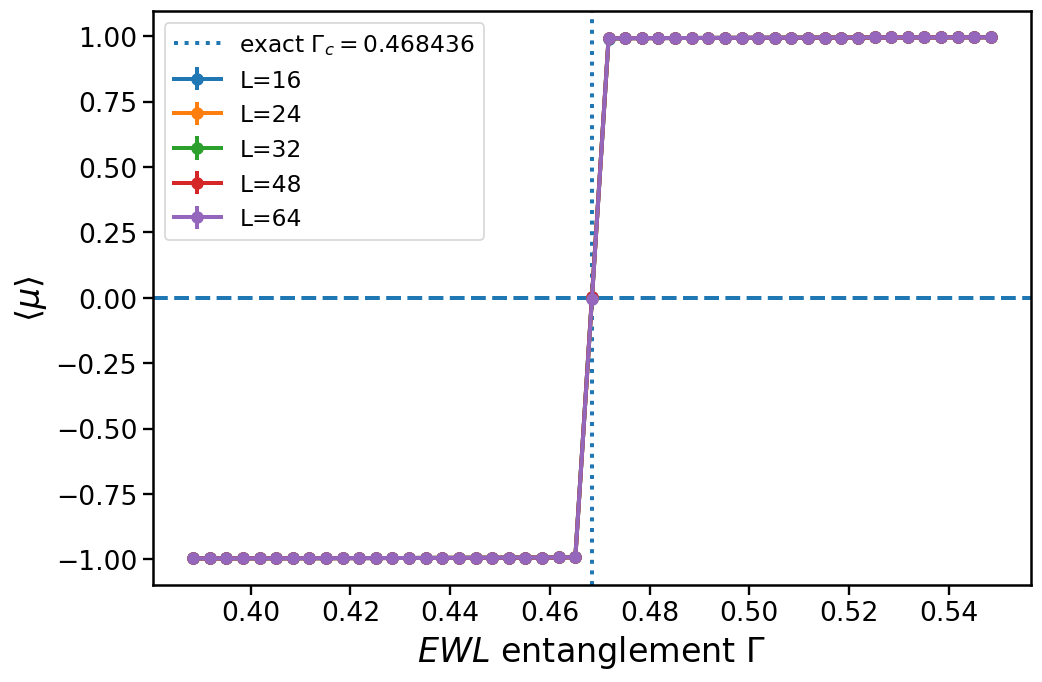}
    \caption{}
    \label{fig6c}
\end{subfigure}
\hfill
\begin{subfigure}[t]{0.45\textwidth}
    \centering
    \includegraphics[width=\linewidth]{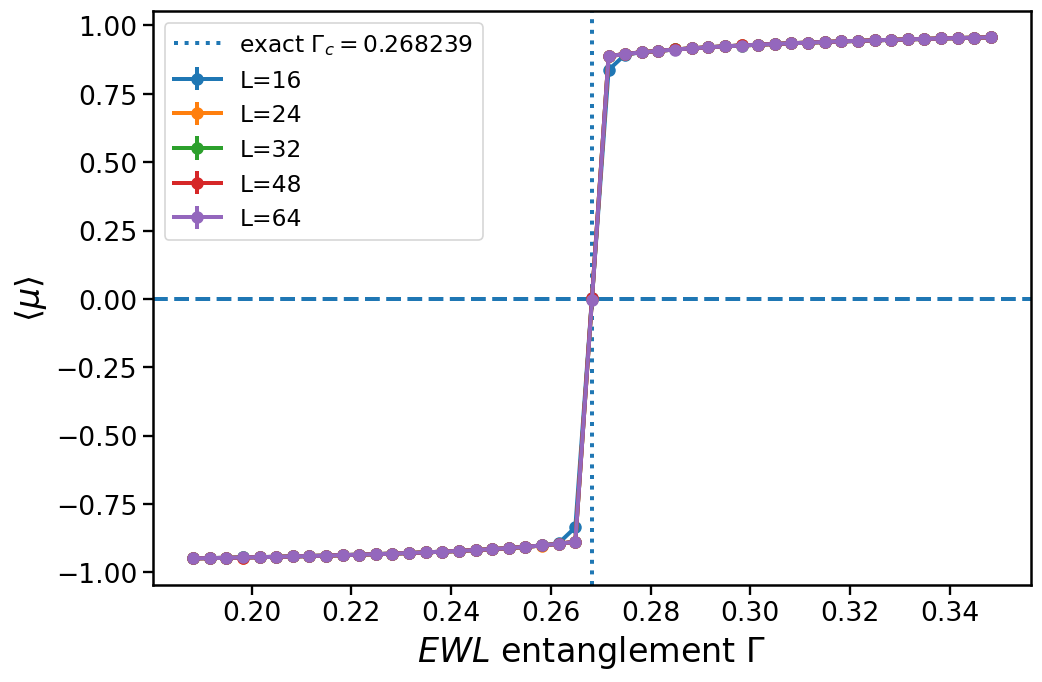}
    \caption{}
    \label{fig6d}
\end{subfigure}

\caption{Amplitude damping channel location: \textbf{(a)} Exact $\mathfrak{H}=0$ branches for resource and post-strategy damping. The post-strategy branch terminates at $p=1/3$, whereas the resource branch remains present over the full range shown. \textbf{(b)} Corresponding $\kappa = \beta \mathfrak{J}_c$ at $\beta=1$; the horizontal line marks the square-lattice critical value $\kappa_c$. \textbf{(c,d)} $\langle \mu \rangle$ across nonzero-field $\Gamma$ scans at $p=0.20$ for resource and post-strategy damping, respectively. The vertical dotted lines indicate the exact $\mathfrak{H}=0$ points: $\mathfrak{H}<0$ favors $\mathsf{D}$ to the left, while $\mathfrak{H}>0$ favors $\mathsf{Q}$ to the right. The shift in the switching point directly shows the effect of moving the amplitude damping channel across the strategic layer.}
\label{fig6}
\end{figure*}

\subsection{\label{subsec5c}Amplitude damping: importance of channel position}
Unlike phase damping (which we do \textit{not} discuss here explicitly) and depolarization, amplitude damping is not covariant with respect to the $\mathsf{Q/D}$ strategic layer. Here, moving the same channel from the \textit{resource} to the \textit{post-strategy} position changes both the location and nature (topology) of the neutrality plot.

At $p=0.20$ and $\beta=1$, the resource-damping branch has,
\begin{equation}
  \Gamma_c^{(\mathcal{R})}=0.468436,~~
  \mathfrak{J}_c^{(\mathcal{R})}=0.724146,
  \label{eq75}
\end{equation}
whereas post-strategy damping gives us,
\begin{equation}
  \Gamma_c^{(\mathcal{S})}=0.268239,~~
  \mathfrak{J}_c^{(\mathcal{S})}=0.480000.
  \label{eq76}
\end{equation}
Both values satisfy $\beta \mathfrak{J}>\kappa_c$, and both therefore show coexistence scaling. The placement effect is visible directly in the nonzero-field magnetization scans in Figs.~\ref{fig6c}, \ref{fig6d}: moving the same amplitude-damping channel shifts the $\mathsf{D}\leftrightarrow \mathsf{Q}$ reversal from $\Gamma_c^{(\mathcal{R})}\simeq 0.4684$ to $\Gamma_c^{(\mathsf{S})}\simeq 0.2682$. As in the depolarizing scans, $\Gamma<\Gamma_c$ lies on the $\mathfrak{H}<0$ ($\mathsf{D}$-favoring) side and $\Gamma>\Gamma_c$ on the $\mathfrak{H}>0$ ($\mathsf{Q}$-favoring) side.

The corresponding susceptibility peak finite-size fits give us,
\begin{equation}
  1.99993\pm0.00004~\text{(\textit{resource})},
~
  1.9960\pm0.0011~\text{(\textit{post-strategy})},
  \label{eq77}
\end{equation}
consistent with $L^2$; the supporting log--log plots are given in Appendix~\ref{appendixF}. These small statistical errors should not be interpreted as a precise determination of a new \textit{exponent}; the relevant conclusion is consistency with the expected two-phase volume scaling.

What is important to notice is the exact branch structure in Fig.~\ref{fig6a}. The post-strategy neutrality line reaches $\Gamma=0$ at $p=1/3$ and disappears for larger $p$, while resource damping shifts the neutrality point toward larger entanglement and retains a positive $\mathfrak{J}$ throughout. At $\beta=1$, the post-strategy coupling has the same $(1-p)^2$ factor as the depolarizing channel, so its ordered coexistence segment terminates at the same Ising condition $p_{*}$ before the microscopic $\mathfrak{H} = 0$ branch itself disappears at $p=1/3$.

\subsection{\label{subsec5d}Relation to previous quantum game and thermodynamic limit works}
Several techniques and methods used in this article have appeared separately in earlier studies, but their combination here differs in several technically important ways.

First, noisy \textit{EWL} games have been studied extensively in finite-player settings, where physical quantum channels modify payoff landscapes and, in some cases, the corresponding equilibrium structure~\cite{ref18, ref19, ref20, ref21, ref22}. Banu and Rao's EDNE analysis, for example, considers two-player EWL games and follows noise and entanglement dependent thresholds at which Nash equilibria change, while noisy multiplayer studies have examined finite multi-qubit games and the robustness of quantum advantage~\cite{ref23, ref24}. Recent \textit{EWL} multiplayer studies have further identified symmetric Nash equilibria in the quantum Volunteer's dilemma and quantified how their existence depends on the entanglement strength and player number~\cite{ref44, ref45}. Separately, Thomas and Balakrishnan introduced a Hamiltonian-driven modified-\textit{EWL} construction based on transverse-field Ising dynamics, including a Stag-Hunt analysis, and subsequently examined amplitude damping decoherence in that framework~\cite{ref46, ref47}. Huang and Qiu consider noisy evolution in different segments of the \textit{EWL} circuit surrounding the strategic operations~\cite{ref20}. These works concern finite games and payoff/equilibrium properties of their quantum resources. In this work, Table~\ref{table1} instead compares two controlled \emph{single-insertion} placements. The resulting noisy payoff matrix is then carried into an exact pair potential and used to determine whether the corresponding short range \textit{2D} Gibbs population is ordered, critical, or disordered.

Second, spatial quantum games on lattices and networks have been studied using evolutionary and cellular automaton dynamics~\cite{ref28, ref29, ref30}. The recent graph-encoded framework of Tsakiroglou \emph{et al.}~\cite{ref31} is particularly close in its edge-based organization, since each graph edge carries a separate two-player quantum game. The dynamical setting is nevertheless different -- the strategies are edge dependent and evolve over repeated rounds through EXP3, the simulations use ideal state-vector dynamics, and decoherence is left as a future extension. Eq.~\eqref{eq34}, by contrast, defines an equilibrium potential game ensemble with a single strategic variable at each site, while the physical channel is already considered into the exact bond parameters $\mathfrak{J}_{\varepsilon}$ and $\mathfrak{H}_{\varepsilon}$. The two constructions therefore describe different dynamical and equilibrium settings and need not lead to the same stationary or critical behavior~\cite{ref6, ref7}.

Third, the earlier construction of Ref.~\cite{ref14} used a \textit{1D} site-based picture in which two locally entangled quantum players occupied each Ising site, while neighboring sites were connected through a separate classical coupling. In the present construction, there is instead one classical strategic variable per site, and the quantum \textit{EWL} game itself is assigned to each nearest-neighbor bond. Hence, the noisy quantum calculation directly generates the lattice coupling $\mathfrak{J}$. The $\mathsf{Q/D}$ game parametrization considered in Ref.~\cite{ref14} also gave $\mathfrak{J}=0$, whereas the Stag-Hunt payoff used here produces a nonzero ferromagnetic interaction and can therefore support a genuine short-range ordered transition. The related classical NEM/ABM study of Ref.~\cite{ref11} provided a payoff-difference mapping and numerical example, but within a \textit{1D} Ising framework and without physical Kraus decoherence.

A separate conceptual issue in quantum game theory is classical reproducibility once an effective payoff table has been specified~\cite{ref48}. The present construction does not claim an intrinsically quantum many-body Gibbs phase: after the noisy two-qubit calculation has produced $\bar{\Delta}_{\varepsilon}$ and hence $(\mathfrak{J}_{\varepsilon},\mathfrak{H}_{\varepsilon})$, the extended lattice is an ordinary classical Ising model. The quantum mechanical content lies in how the open-system \textit{EWL} circuit generates the \textit{channel} and \textit{placement} dependent effective interaction.

Finally, Ref.~\cite{ref32} develops a complementary potential game description of phase transition analogies, including quantum strategies, primarily for highly connected populations. The present setting differs in two important respects. Here $p$ represents the strength of a physical open-system channel acting inside the \textit{EWL} circuit and is kept distinct from the strategic noise scale $\beta^{-1}$. In addition, the collective transition considered here is the finite $\beta$ critical point of a N.N. square lattice. This distinction is also why Sec.~\ref{subsec2f} separates the two-player Nash conditions from the thermodynamic neutrality and criticality conditions.

The resulting work can be summarized as,
\begin{equation}
  \varepsilon_p
  \to
  \bar{\Delta}_{\varepsilon}
  \to
  (\mathfrak{J}_{\varepsilon},\mathfrak{H}_{\varepsilon})
  \to
  \text{short-range 2D phase structure}.
\end{equation}

Physical decoherence and strategic noise therefore remain separate control parameters throughout the construction. To the best of current knowledge, the noisy finite-player, multiplayer, graph-based, and thermodynamic game studies do not carry this full sequence through for a physical channel acting inside the \textit{EWL} circuit. A direct consequence is that microscopic neutrality and collective interaction strength need not vary together: two channels may share the same $\mathfrak{H}=0$ point while placing the population in different thermodynamic regimes.

\subsection{\label{subsec5e}Neutrality \& role of entanglement}
The results also separate microscopic strategic neutrality from collective criticality. In an exact potential game, $\mathfrak{H}=0$ only means that the two homogeneous Ising orientations are equally favored by the potential. Whether the population is ordered is determined independently by the value of $\beta \mathfrak{J}$~\cite{ref1, ref2}. Consequently, two channels can have the same neutrality curve $\Gamma_c(p)$ while lying in different thermodynamic regimes.

For the current payoff matrix, the unentangled limit provides a further distinction. At $\Gamma=0$,
\begin{align}
 \mathfrak{H}_{ph}(0,p)&=-\frac14,\\
 \mathfrak{H}_{ dep}(0,p)&=-\frac{1-p}{4},\\
 \mathfrak{H}_{ad}^{(\mathcal{R})}(0,p)&=-\frac{1}{4},
 \label{eq78}
\end{align}
so none of these interacting channel families reaches $\mathfrak{H}=0$ at \textit{zero} entanglement. The depolarizing endpoint $p=1$ is the trivial limit $\mathfrak{J}=\mathfrak{H}=0$. Post-strategy amplitude damping is different,
\begin{equation}
 \mathfrak{H}_{ ad}^{(\mathcal{S} )}(0,p)=\frac{(1-p)(3p-1)}{4},
 \label{eq79}
\end{equation}
which vanishes at $p=\frac{1}{3}$. This is exactly the point at which the positive entanglement neutrality branch in Eq.~\eqref{eq45} reaches $\Gamma=0$ and terminates. Thus, within the channel families considered here and the restricted $\mathsf{Q/D}$ strategy set, all nontrivial neutrality branches require a nonzero \textit{EWL} entangling parameter $\Gamma$, apart from the post-strategy amplitude damping endpoint at $p = \frac{1}{3}$. Indeed, along the depolarizing neutrality branch the resource negativity vanishes at $p\simeq0.280719$ while the interacting
$\mathfrak{H}=0$ branch persists.

\section{Conclusion}
The present work establishes a direct link between physical decoherence in a microscopic \textit{EWL} game and equilibrium collective behavior in a short-range \textit{2D} population. The quantum calculation is performed independently for each N.N. interaction, producing a channel and placement dependent payoff matrix. This payoff matrix determines the classical pair interaction $\mathfrak{J}$ and field $\mathfrak{H}$, which are then used directly in the square-lattice Gibbs model without additional phenomenological fitting.

The main result is that decoherence can do more than modify a two-player payoff landscape: it can renormalize the effective interaction and thereby place the population in different thermodynamic regimes. Phase damping and depolarization provide the clearest example. They share the same microscopic neutrality line $\mathfrak{H}=0$, but only depolarization weakens the coupling as $(1-p)^2$. At $\beta=1$, the depolarizing branch therefore reaches the exact critical point $p_{*}\approx 0.233460\ldots$, where the field-driven coexistence line terminates. Zero-field Binder crossings and finite-size scaling are consistent with the expected \textit{2D} Ising critical behavior. The nonzero-field scans show what this critical point separates: for $p<p_{*}$, crossing $\mathfrak{H}=0$ produces two-phase coexistence, with $\chi_\mu^{\rm peak}\sim L^2$ and a bimodal $P(\mu)$, whereas for $p>p_{*}$ the same change in the sign of $\mathfrak{H}$ produces only a smooth crossover with finite susceptibility.

Amplitude damping provides a separate example in which the circuit placement of the channel matters. Resource damping preserves an interacting zero-field branch throughout the physical noise interval, whereas post-strategy damping shifts the branch toward zero entanglement and removes it beyond $p=\frac{1}{3}$. This difference follows from the non-unital and non-Pauli-covariant character of amplitude damping rather than from the numerical implementation.

Several limitations define the scope of these results. The extended lattice is an ordinary classical Ising system rather than a globally entangled quantum lattice; quantum mechanics enters locally through the noisy \textit{EWL} encounters that generate the effective bond potential. The equilibrium Gibbs construction describes a potential-game population with logit-type strategic fluctuations and is not intended as a universal model of evolutionary dynamics. The analysis is also restricted to a single Stag-Hunt payoff matrix and the $\mathsf{Q/D}$ strategy set. Extending the channel-to-interaction mapping to more general games, larger strategy spaces, correlated noise, and non-equilibrium spatial dynamics provides a natural direction for further study.

\section*{ACKNOWLEDGMENTS}
The author acknowledges the use of GPT-5.6 Sol as a supporting tool during the preparation of this manuscript, primarily for assistance with grammar, proofreading, code drafting and debugging (as disclosed in Sec.~\ref{sec4}), and LaTeX formatting. The underlying physical ideas and theoretical framework were developed by the author, and all calculations and computational results were performed, reviewed, and verified by the author.

\section*{Data \& code availability}
The numerical data and source code related to the figures and findings of this article are available from the author upon reasonable request. No external datasets were used.

\appendix

\section{\label{appendixA}Noiseless \textit{EWL} derivation}
In this appendix, we show the state-vector calculation behind Eq.~\eqref{eq13}. We define,
\begin{equation}
 c=\cos\frac{\Gamma}{2},
 ~~
 s=\sin\frac{\Gamma}{2},
 ~~
 |{\Omega_1}\rangle=c|{00}\rangle+is|{11}\rangle.
\end{equation}
From Eq.~\eqref{eq5},
\begin{align}
 \mathsf{Q}|0\rangle =i|0\rangle,~~ & \mathsf{Q} |1\rangle =-i|1\rangle,\\
 \mathsf{D}|0\rangle =-|1\rangle,~~& \mathsf{D}|1\rangle =|0\rangle.
\end{align}
The inverse entangler is,
\begin{equation}
 \mathcal{J}^{\dagger}(\Gamma) = c\,\mathbb{I}\otimes \mathbb{I} +is\,\mathbb{Y}\otimes \mathbb{Y}.
\end{equation}
For the $\mathsf{QQ}$ strategy pair,
\begin{equation}
 (\mathsf{Q}\otimes \mathsf{Q})|\Omega_1\rangle =
 -c|00\rangle-is |11\rangle,
\end{equation}
and application of $\mathcal{J}^{\dagger}(\Gamma)$ gives us $-|00\rangle$. Hence, the row-player payoff is $\mathsf{R}$.

For $\mathsf{QD}$ case,
\begin{equation}
 (\mathsf{Q}\otimes \mathsf{D})|\Omega_1\rangle = -ic|01\rangle+s |10\rangle,
\end{equation}
and therefore,
\begin{equation}
 |\Omega_{\mathsf{QD}}\rangle
 = -i\cos\Gamma\,|01\rangle +\sin\Gamma\,|{10}\rangle.
\end{equation}
The two nonzero outcome probabilities are,
\begin{equation}
 P_{01}=\cos^2\Gamma,~~ P_{10}=\sin^2\Gamma,
\end{equation}
which give the row-player payoff as,
\begin{equation}
 \mathsf{S}\cos^2\Gamma+\mathsf{T}\sin^2\Gamma.
\end{equation}
Similarly,
\begin{equation}
 |{\Omega_{\mathsf{DQ}}}\rangle = \sin\Gamma\,|{01}\rangle -i\cos\Gamma\,|{10}\rangle,
\end{equation}
so that the row-player payoff is,
\begin{equation}
 \mathsf{T}\cos^2\Gamma+\mathsf{S}\sin^2\Gamma.
\end{equation}
Finally, the $\mathsf{DD}$ strategy pair gives $|{11}\rangle$ and hence the payoff $\mathsf{P}$. From these four entries, we get Eq.~\eqref{eq13}. The calculation also shows explicitly that the nontrivial $\Gamma$ dependence of the reduced $\mathsf{Q/D}$ game appears in the off-diagonal strategic outcomes.

\section{\label{appendixB}Full noisy payoff matrices}
Here, we calculate the complete noisy \textit{row}-player payoff matrices from which the effective parameters in Table~\ref{table1} are determined.

For phase damping, the resource and post-strategy placements give the same payoff matrix,
\begin{equation}
 \bar{\Delta}_{ ph} = \begin{bmatrix}
 4-p\sin^2\Gamma & \dfrac{3}{2}(2-p)\sin^2\Gamma
 \\[2mm]
 3-\dfrac{3}{2}(2-p)\sin^2\Gamma & 2+p\sin^2\Gamma
 \end{bmatrix}.
 \label{eqB1}
\end{equation}
For depolarization, the two placements are again equivalent. Hence, we have,
\begin{equation}
 \bar{\Delta}_{dep} = \begin{pmatrix}
 m & n\\
 r & q
\end{pmatrix},
\end{equation}
with
\begin{gather}
 m = 4-\frac52p+\frac34p^2 -p(1-p)\sin^2\Gamma,
 \\
 n = 3p-\frac34p^2 +\frac32(1-p)(2-p)\sin^2\Gamma,
 \\
 r = 3-\frac34p^2 -\frac32(1-p)(2-p)\sin^2\Gamma,
 \\
 q = 2-\frac12p+\frac34p^2 +p(1-p)\sin^2\Gamma.
\end{gather}
At $p=1$, all four entries reduce to $9/4$, as expected for completely \textit{mixed} local qubits under the depolarizing convention of Eq.~\eqref{eq18}.

For amplitude damping, it is convenient to define
\begin{equation}
 x=\sin^2\frac{\Gamma}{2}.
\end{equation}
Hence, for resource damping,
\begin{equation}
 \bar{\Delta}_{ ad}^{(\mathcal{R})}
 =
 \begin{bmatrix}
 m' & n' \\[1mm]
 r' & q' 
 \end{bmatrix}.
 \label{eqB8}
\end{equation}
where, we have,
\begin{gather}
    m' = 4+(3p^2-5p)x\\
    n' = 3x[-p^2+4px-2p-4x+4]\\
    r' = -3[p^2x+4px^2-4px-4x^2+4x-1]\\
    q' =  2+(3p^2-p)x.
\end{gather}
For post-strategy damping,
\begin{equation}
 \bar{\Delta}_{ ad}^{(\mathcal{S})}
 =
 \begin{bmatrix}
 m'' & n'' \\[1mm]
 r'' & q'' 
 \end{bmatrix}.
 \label{eqB13}
\end{equation}
where, we have,
\begin{gather}
    m'' = 4+(3p^2-5p)x\\
    n'' = 2(6px^2-7px+2p-6x^2+6x)\\
    r'' = -12px^2+10px+p+12x^2-12x+3 \\
    q'' =   -3p^2x+3p^2+px-p+2.
\end{gather}
By substituting the above amplitude damping payoff entries into Eq.~\eqref{eq27}, we get the channel-dependent $\mathfrak{J}$ and $\mathfrak{H}$ as given in Table~\ref{table1}.

\section{\label{appendixC}Resource density matrices and negativity}
The microscopic entanglement expressions used in the main text are evaluated for the two-qubit \textit{EWL} resource before the strategy operations. We define,
\begin{equation}
 c=\cos\frac{\Gamma}{2},
 ~~
 s=\sin\frac{\Gamma}{2}.
\end{equation} 
The noiseless resource density matrix is given as,
\begin{equation}
 \varrho_1=
 \begin{bmatrix}
 c^2&0&0&-ics\\
 0&0&0&0\\
 0&0&0&0\\
 ics&0&0&s^2
 \end{bmatrix}.
 \label{eqC2}
\end{equation}
Under independent phase damping on the two qubits,
\begin{equation}
 \varrho_{ ph}^{(\mathcal{R})}
 =
 \begin{bmatrix}
 c^2&0&0&-i(1-p)cs\\
 0&0&0&0\\
 0&0&0&0\\
 i(1-p)cs&0&0&s^2
 \end{bmatrix}.
 \label{eqC3}
\end{equation}
After partial transpose, the only potentially negative eigenvalue is,
\begin{equation}
 -(1-p)cs,
\end{equation}
which gives us,
\begin{equation}
 \mathcal{N}_{ ph}(\Gamma,p)
 =
 \frac{1-p}{2}\sin\Gamma.
\end{equation}
For resource amplitude damping,
\begin{equation}
 \varrho_{ ad}^{(\mathcal{R})}
 =
 \begin{bmatrix}
 c^2+p^2s^2&0&0&-i(1-p)cs\\
 0&p(1-p)s^2&0&0\\
 0&0&p(1-p)s^2&0\\
 i(1-p)cs&0&0&(1-p)^2s^2
 \end{bmatrix}.
 \label{eqC6}
\end{equation}
The central $2\times2$ block of the partial transpose has eigenvalues,
\begin{equation}
 p(1-p)s^2\pm(1-p)cs.
\end{equation}
The negative eigenvalue, when present, therefore gives us,
\begin{align}
 \mathcal{N}_{ ad}(\Gamma,p)
  = (1-p)\max\!\left(0,~cs-ps^2\right)
 \nonumber\\
 = \frac{1-p}{2} \max\!\left[ 0,\, \sin\Gamma-p(1-\cos\Gamma) \right].
 \label{eqC8}
\end{align}
Over the physical parameter interval considered in the main text, the quantity inside the maximum is nonnegative, giving the form used in Eq.~\eqref{eq59c}.

For depolarization, the linear extension of Eq.~\eqref{eq18} to an arbitrary two-qubit operator gives us,
\begin{equation}
 (\mathcal{D}_p^{\otimes2})(\varrho)
 = (1-p)^2\varrho + \frac{p(1-p)}{2} \left[
 \mathbb{I}\otimes\varrho_B+\varrho_A\otimes \mathbb{I} \right] + \frac{p^2}{4}\mathbb{I}\otimes \mathbb{I},
 \label{eqC9}
\end{equation}
where $\varrho_A$ and $\varrho_B$ are the reduced density matrices. By applying Eq.~\eqref{eqC9} to
Eq.~\eqref{eqC2} and diagonalizing the partial transpose, we get,
\begin{equation}
 \mathcal{N}_{ dep}(\Gamma,p) =
 \max\!\left[0,\, \frac{(1-p)^2}{2}\sin\Gamma -\frac{p(2-p)}{4} \right].
\end{equation}
These calculations use the standard negativity definition of Ref.~\cite{ref41} and concern only the microscopic two-qubit resource. They are independent of the classical lattice sampling.

\section{\label{appendixD}Independent density-matrix \& algebraic validation}
The analytical channel-to-interaction expressions were checked using an independent $4\times4$ density-matrix implementation. For randomly selected $(\Gamma,p)$ points, the numerical routine constructs $\mathcal{J}(\Gamma)$, applies the appropriate Kraus map at the specified circuit position, evaluates the four $\mathsf{Q/D}$ row-player payoffs from Eq.~\eqref{eq10}, and obtains
$\mathfrak{J}$ and $\mathfrak{H}$ from Eq.~\eqref{eq27}. The same calculation independently
diagonalizes the partial transpose of the resource density matrix.

The largest absolute discrepancies observed in the numerical checks are given in Table~\ref{table-app1}. The discrepancies are at the level of ordinary double-precision roundoff. These checks establish numerical consistency with the closed form expressions and are independent of the Monte Carlo error analysis.

\begin{center}
\begin{table*}
\begin{tabular}{lccc}
\toprule
channel
& $\max|\Delta\mathfrak{J}|$
& $\max|\Delta\mathfrak{H}|$
& $\max|\Delta\mathcal{N}|$
\\
\midrule
noiseless
& $1.11\times10^{-16}$
& $2.22\times10^{-16}$
& $5.55\times10^{-17}$
\\
phase damping
& $2.22\times10^{-16}$
& $3.89\times10^{-16}$
& $1.11\times10^{-16}$
\\
depolarizing
& $3.33\times10^{-16}$
& $4.44\times10^{-16}$
& $1.39\times10^{-16}$
\\
amplitude, resource
& $2.22\times10^{-16}$
& $3.61\times10^{-16}$
& $1.11\times10^{-16}$
\\
amplitude, post-strategy
& $2.78\times10^{-16}$
& $3.33\times10^{-16}$
& $5.55\times10^{-17}$
\\
\bottomrule
\end{tabular}
\caption{Discrepancy values observed in numerical checks.}
\label{table-app1}

\end{table*}
\end{center}

\section{\label{appendixE}Additional finite-size and convergence diagnostics}
The depolarizing zero-field scan uses four independent replicas and $8000$ recorded measurements per replica after subsampling. As expected for local Metropolis dynamics near a continuous transition, critical slowing down is visible in the integrated autocorrelation time
~\cite{ref39, ref43}. For each replica we define,
\begin{equation}
\tau_{\max} =
\max\left\{
\tau_{\rm int}^{(|\mu|)},
\tau_{\rm int}^{(\mu^2)},
\tau_{\rm int}^{(E_{\rm lat})}
\right\}.
\end{equation}
where, $E_{\rm lat}\equiv \mathsf H(\{\mathfrak{s}_i\})$. The largest measured $\tau_{\max}$ and the smallest effective sample count obtained from $|\mu|$ at each lattice size $L$ are,
\begin{center}
\begin{tabular}{ccc}
\toprule
$L$
& $\tau_{\max}$
& minimum $n_{\rm eff}(|\mu|)$
\\
\midrule
16 & 2.40 & 1666.8\\
24 & 5.77 & 693.4\\
32 & 11.75 & 340.4\\
48 & 39.52 & 101.2\\
64 & 48.93 & 81.8\\
\bottomrule
\end{tabular}
\end{center}
Here the autocorrelation times are measured in units of recorded samples; consecutive recorded configurations are separated by \textit{five} Metropolis sweeps.

For the dedicated simulations at the exact $p_{*}$, the measurement window is increased to $100000$ sweeps per replica. The corresponding maximum autocorrelation times are,
\begin{equation}
 2.15,\;4.69,\;8.59,\;22.86,\;38.11
\end{equation}
for $L=16,24,32,48,64$, respectively. The minimum effective sample counts obtained from $|\mu|$ are
\begin{equation}
 4653,\;2133,\;1164,\;437,\;262.
\end{equation}
The reported uncertainties of the thermodynamic observables are obtained from the independent replicas. The autocorrelation-time and blocking estimates are used as convergence diagnostics for the individual Monte Carlo trajectories.

\section{\label{appendixF}Additional plots}

\begin{figure*}
\centering
\includegraphics[width=0.48\linewidth]
{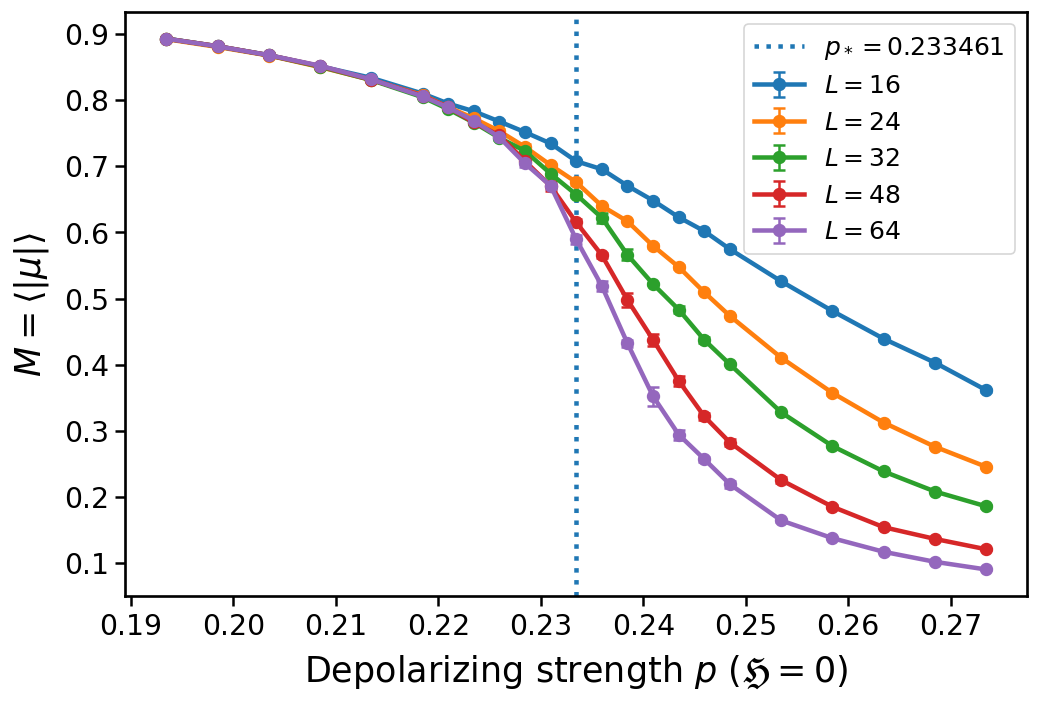}
\includegraphics[width=0.48\linewidth]
{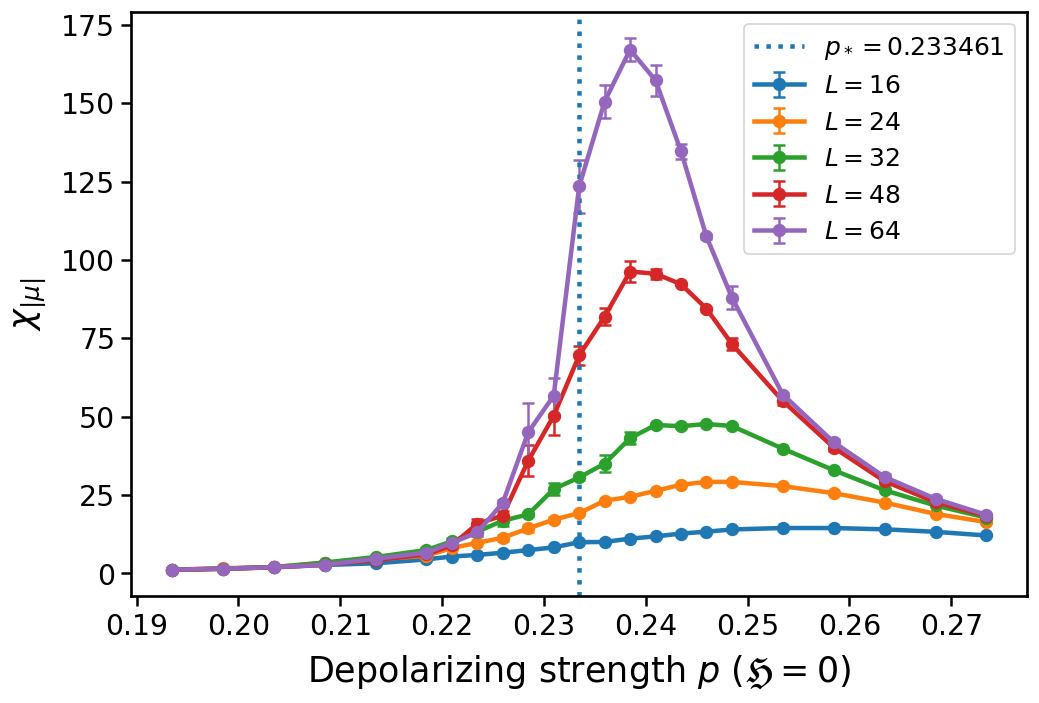}
\caption{Additional depolarizing zero-field data at $\beta=1$. \textbf{(a)} $M=\langle|\mu|\rangle$ and \textbf{(b)} $\chi_{|\mu|}$ as functions of $p$ along the exact $\mathfrak{H}=0$ line.}
\label{figapp1}
\end{figure*}

\begin{figure*}
\centering
\includegraphics[width=\linewidth]{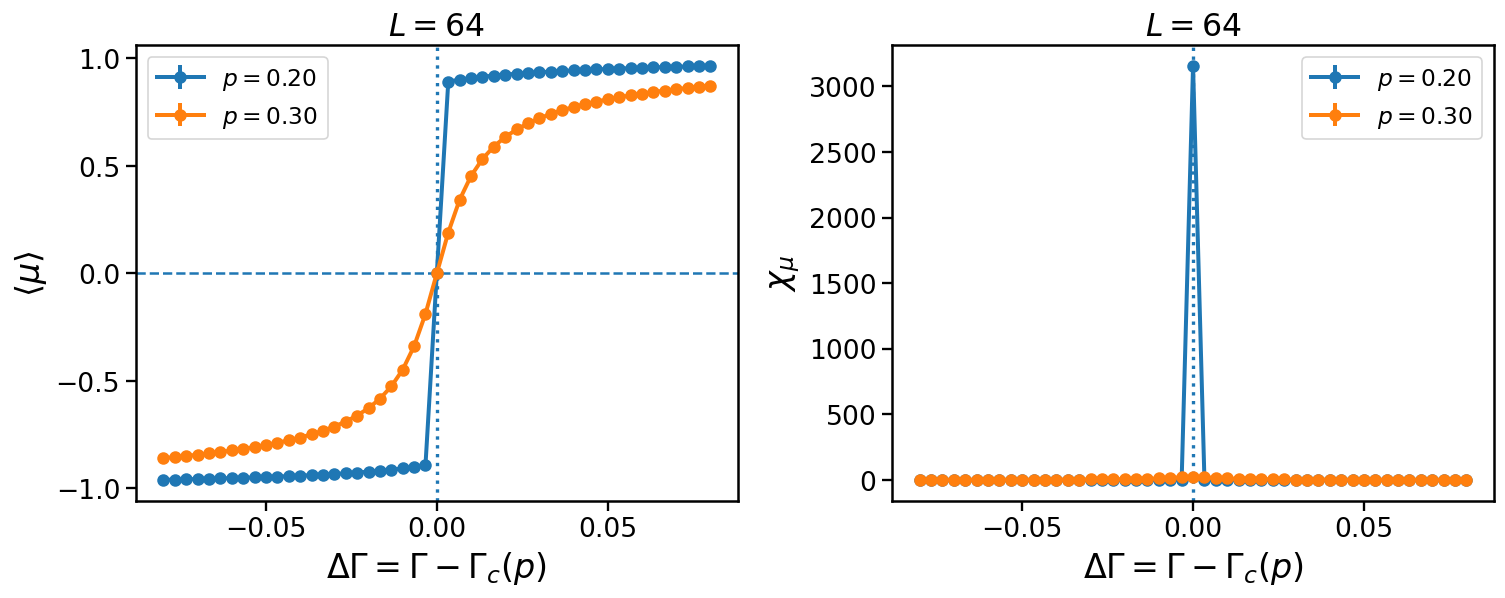}
\caption{Comparison of the depolarizing field scans at $L=64$ and $\beta=1$. The horizontal coordinate is measured relative to the exact neutrality point, $\Delta\Gamma=\Gamma-\Gamma_c(p)$, such that $\Delta\Gamma=0$ corresponds to $\mathfrak{H}=0$. \textbf{(a)} Signed magnetization for $p=0.20<p_{*}$ and $p=0.30>p_{*}$. The ordered case shows an abrupt reversal between the $\mathsf{D}$ and $\mathsf{Q}$ dominated states, whereas the disordered case changes smoothly through zero. \textbf{(b)} Corresponding signed susceptibility $\chi_\mu$. The large peak for $p=0.20$ reflects two-phase coexistence, while the response for $p=0.30$ remains finite.}
\label{figapp3}
\end{figure*}

\begin{figure*}
\centering
\includegraphics[width=0.48\linewidth]
{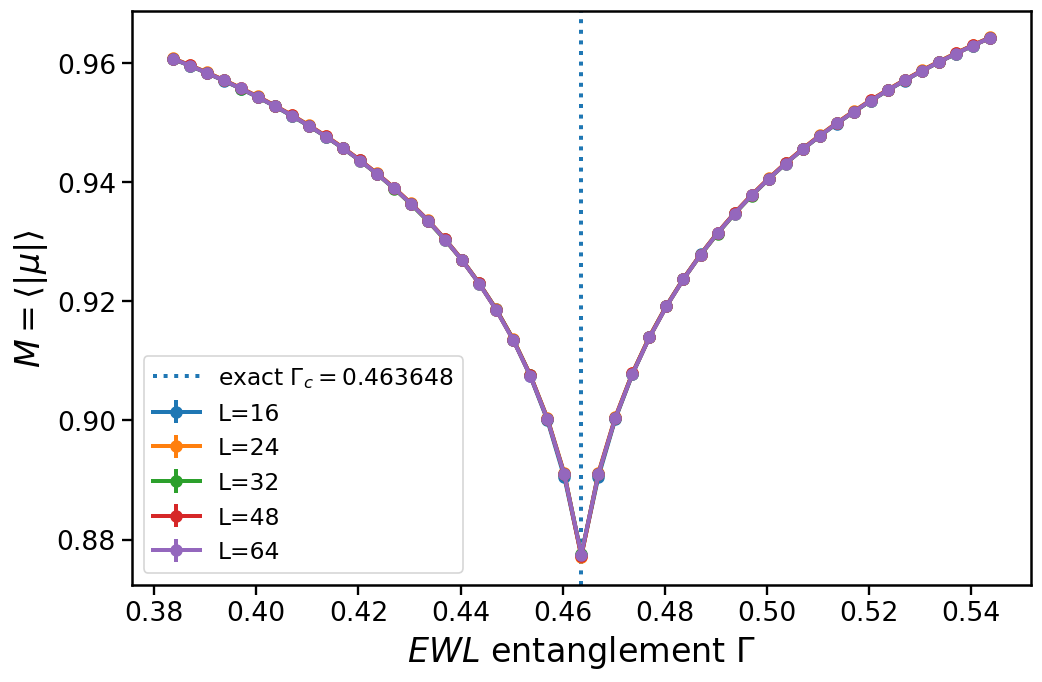}
\includegraphics[width=0.48\linewidth]
{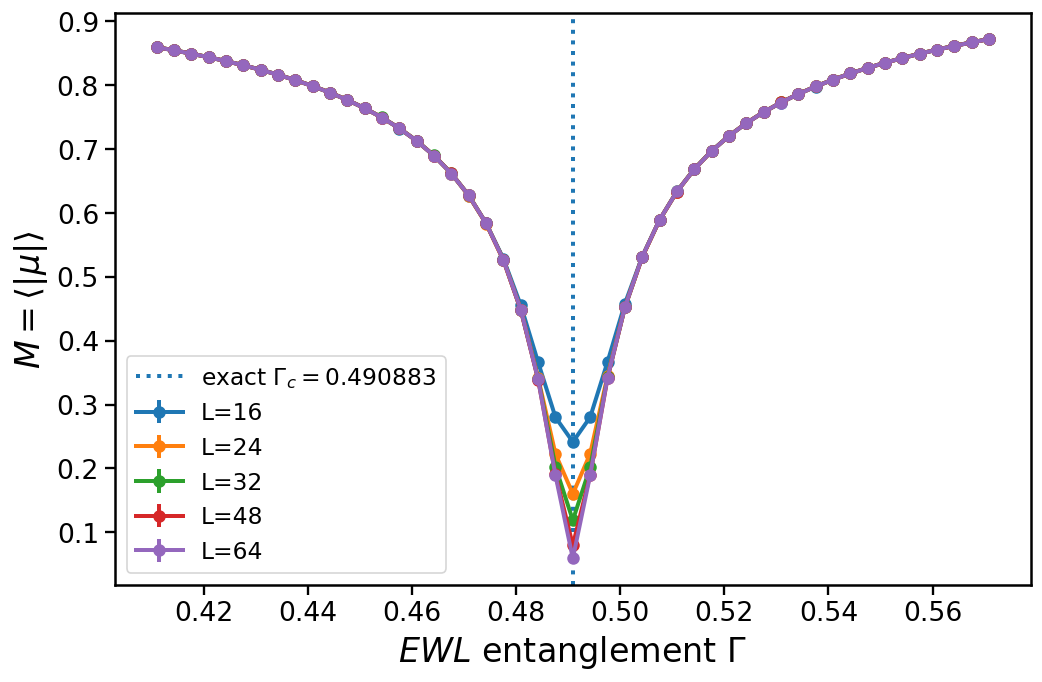}
\caption{Finite-size behavior of the absolute magnetization $M=\langle|\mu|\rangle$ across the depolarizing neutrality point at $\beta=1$. \textbf{(a)} For $p=0.20<p_{*}$, the minimum at the exact $\mathfrak{H}=0$ point remains finite and nearly independent of $L$, consistent with an ordered coexistence point. \textbf{(b)} For $p=0.30>p_{*}$, the value of $M$ at $\mathfrak{H}=0$ decreases systematically with increasing $L$, consistent with the disordered regime. The vertical dotted lines mark the corresponding exact values of $\Gamma_c(p)$.}
\label{figapp4}
\end{figure*}

\begin{figure*}
\centering
\includegraphics[width=0.48\linewidth]
{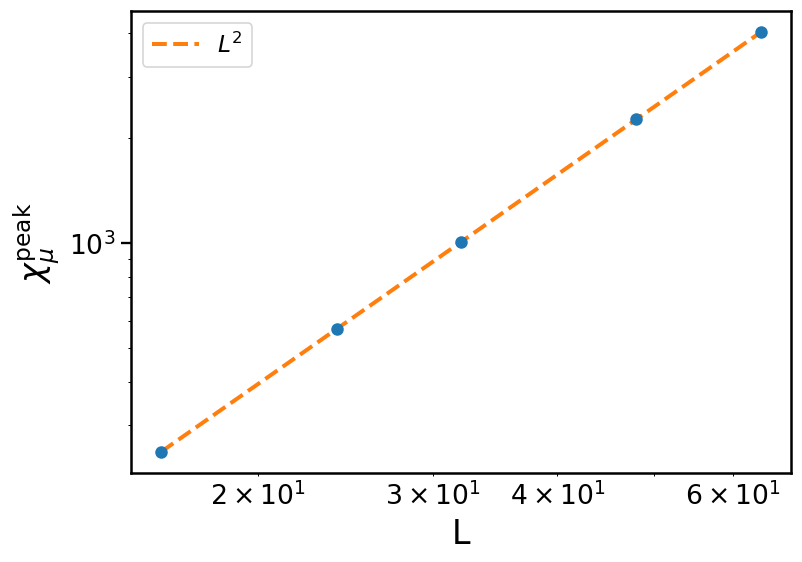}
\includegraphics[width=0.48\linewidth]
{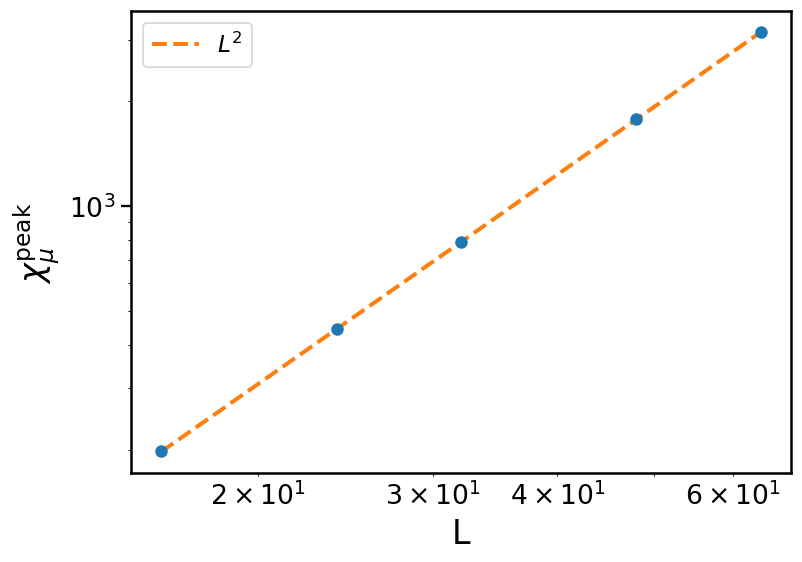}
\caption{Susceptibility peak finite-size scaling at $p=0.20$ for \textbf{(a)} resource amplitude damping and \textbf{(b)} post-strategy amplitude damping. Both are consistent with the expected
$\chi_{\mu}^{\rm peak}\sim L^2$ volume scaling of two-phase coexistence, complementing the signed nonzero-field switching curves in Figs.~\ref{fig6c}, \ref{fig6d}.}
\label{figapp2}
\end{figure*}

\end{document}